\documentclass[12pt]{article}

\usepackage{a4}
\usepackage{pslatex}
\usepackage{hyperref}
\usepackage[latin1]{inputenc}
\usepackage[T1]{fontenc}
\usepackage{graphicx}
\usepackage{amsfonts}
\usepackage{latexsym}
\usepackage{color,pstcol}
\usepackage{colortbl}

\makeatletter
\def\@sect#1#2#3#4#5#6[#7]#8{\ifnum #2>\c@secnumdepth
  \def\@svsec{}\else 
  \refstepcounter{#1}\edef\@svsec{\csname the#1\endcsname.\hskip0.5em}\fi
  \@tempskipa #5\relax
  \ifdim \@tempskipa>\z@
  \begingroup 
     #6\relax
     \@hangfrom{\hskip #3\relax\@svsec}{\interlinepenalty \@M #8\par}%
  \endgroup
  \csname #1mark\endcsname{#7}\addcontentsline
      {toc}{#1}{\ifnum #2>\c@secnumdepth \else
        \protect\numberline{\csname the#1\endcsname}\fi #7}%
  \else
    \def\@svsechd{#6\hskip #3\@svsec #8\csname #1mark\endcsname
      {#7}\addcontentsline{toc}{#1}{\ifnum #2>\c@secnumdepth \else
        \protect\numberline{\csname the#1\endcsname}\fi #7}}%
  \fi \@xsect{#5}}
\@addtoreset{equation}{section}
\@addtoreset{figure}{section}
\makeatother

\renewcommand\thesection{\Roman{section}}

\renewcommand\theequation{%
  \ifnum \value{section}>0
     \thesection.\arabic{equation}%
  \else
     \arabic{equation}%
  \fi}
\renewcommand\thefigure{%
  \ifnum \value{section}>0
     \thesection.\arabic{figure}%
  \else
     \arabic{figure}%
  \fi}
\def\pt{{p_{\rm T}}}
\def\ptcut{p_{\rm T}^{\rm cut}}
\def\Qb{Q_b}
\def\Qq{Q_q}
\def\R{\right)}
\def\L{\left(}
\def\R{\right)}
\def\Mm{{\cal{M}}}
\def\M{{{\cal M}}}

\def\O{{\rm O}}
\newcommand{\li}{{\rm Li_2}}
\newcommand{\bb}{\bar{b}}
\newcommand{\bbb}{b\bar{b}}
\newcommand{\qqb}{q\bar{q}}
\newcommand{\qb}{\bar{q}}
\newcommand{\ra}{\rightarrow}
\newcommand{\veps}{\varepsilon}
\newcommand{\tf}{t_1}
\newcommand{\ts}{t_2}
\newcommand{\uf}{u_1}
\newcommand{\us}{u_2}

\def\Li2#1{\mbox{Li}_2\left(#1\right)}

\def\kqb{k_{\bar{q}}}
\def\kbb{k_{\bar{b}}}
\def\kb{k_b}
\def\kg{k_{g}}
\def\kq{k_{q}}
\def\e{\varepsilon}

\def\order#1{{\rm O}(#1)}
\def\lra{\leftrightarrow}

\def\muF{{\mu_F}}

\def\nn{\nonumber}
\def\Ref#1{Ref.~\cite{#1}}
\def\Refs#1{Refs.~\cite{#1}}
\def\Eq#1{{Eq.~(\ref{#1})}}

\def\BB{{\rm B}}
\def\m#1{{m_{#1}}}
\def\mt{{m_t}}
\def\mb{{m_b}}

\def\mz{{m_Z}}
\def\mw{{m_W}}
\def\sw{{s_W}}
\def\cw{{c_W}}

\def\gw{{g_W}}

\def\gvq{{g_v^q}}
\def\gaq{{g_a^q}}
\def\gvb{{g_v^b}}
\def\gab{{g_a^b}}
\def\gvt{{g_v^t}}
\def\gat{{g_a^t}}
\def\sumqn{\overline{\sum}}

\def\as{{\alpha_s}}
\def\aas{{\alpha_s^2\alpha}}
\newcommand{\ind}{{\rm d}}

\newcommand{\Amt}{{\overline{\rm A}_0(\mt^2)}}

\newcommand{\Amw}{{\overline{\rm A}_0(\mw^2)}}

\newcommand{\Botmtmw}{\overline{\rm B}_0(t,\mt^2,\mw^2)}
\newcommand{\Bosmtmt}{{\overline{\rm B}_0(s,\mt^2,\mt^2)}}

\newcommand{\DBoomtmw}{{{\partial\over \partial p^2}{\rm B}_0(p^2,\mt^2,\mw^2)\Big|_{p^2=0}}}
\newcommand{\Cotmtmtmw}{{\rm C}_0(t,\mt^2,\mt^2,\mw^2)}
\newcommand{\Cosmtmwmt}{{\rm C}_0(s,\mt^2,\mw^2,\mt^2)}

\newcommand{\Cosmtmtmt}{{\rm C}_0(s,\mt^2,\mt^2,\mt^2)}

\newcommand{\Dotw}{{\rm D}_0^{W}(z)}

\newcommand{\facselfz}{
{2-N^2+N^2z\over N(N^2-1)\*s}(\gvb^2+\gab^2){1+z^2\over(1+z)^2}}
\newcommand{\facselfw}{{2-N^2+N^2z\over
N(N^2-1)\*s}\gw^2{1+z^2\over(1+z)^2}}
\begin{document}
\thispagestyle{empty}
\begin{flushright}
  HU-EP-09/37\\
  SFB/CPP-09-78\\
  TTP09-30
\end{flushright}
\vspace*{3cm}
\begin{center}
  {\Large\bf
    Weak effects in b-jet production at hadron colliders
    }\\
  \vspace*{1cm}

  J.H. K\"uhn$^a$, A. Scharf$^b$, P. Uwer$^c$\\
  \vspace*{0.5cm}
  {\em $^a$Institut f\"ur Theoretische Teilchenphysik,
    Universit\"at Karlsruhe\\ 76128 Karlsruhe, Germany}\\
  {\em $^b$ Department of Physics,  SUNY at Buffalo,\\ Buffalo, NY 14260-1500, USA}\\
  {\em $^c$  Institut f\"ur Physik, Humboldt-Universit\"at zu Berlin,\\  Newtonstr. 15, 12489 Berlin, Germany}\\
\end{center}
\vspace*{1.5cm}
\centerline{\bf Abstract}
\begin{center}
  \parbox{0.8\textwidth}{
    One of the main challenges of the Tevatron at Fermilab and the Large Hadron
    Collider (LHC) at CERN is the
    determination of Standard Model (SM) parameters at the TeV scale.
    In this context various processes will be investigated 
    which involve bottom-quark jets 
    in the final state, for example decays of top-quarks or gauge bosons of the weak interaction.  
    Hence the theoretical understanding of processes with
    bottom-quark jets is necessary. In this paper analytic results 
    will be presented for the weak corrections to bottom-quark
    jet production --- neglecting purely photonic corrections. The
    results will be used to study differential distributions,
    where sizeable effects are observed.}
\end{center}
\newpage

\setcounter{page}{1}
\section{Introduction}
With the start of the LHC a new energy regime is accessible, either to confirm
the Standard Model (SM) or to verify new physics at the TeV scale. Famous possible 
SM extensions are heavy gauge bosons (e.g. $Z'$), Supersymmetry or 
Kaluza-Klein resonances in models with extra dimensions. Beside the outstanding discovery of the Higgs boson, processes involving top-quarks, 
gauge bosons of the weak interaction and jets are of particular interest.
The experimental identification will rely on their characteristic decay products with leptons or bottom jets as characteristic examples.
In particular in processes including top-quarks or a (light) Higgs boson the identification of the bottom-quark jets will play a crucial
role. In addition to these SM processes, bottom-quark jets are also involved in many decays originating from signals for physics beyond the SM. 
In particular new resonances, e.g. heavy gauge bosons, decaying into $b$-jets require a detailed SM-based prediction to prove a deviation from the SM.
The discrimination of $b$-jets from light quark- and gluon-jets, $b$-tagging, makes use of the large lifetime of B-mesons. 
In the last years there were large efforts by several experimental
groups to develop new or to improve established $b$-tagging algorithms. 
This requires a detailed theoretical understanding of the corresponding 
processes like bottom-quark pair and
single bottom-quark production. Both processes were studied in the past. The
differential cross section for bottom-quark pair production is known to
next-to-leading order (NLO) accuracy in quantum chromodynamics (QCD)
\cite{Nason:1989zy,Beenakker:1988bq}. For massless single bottom production the NLO QCD
corrections can be extracted from di-jet production \cite{Ellis:1985er,Giele:1993dj}. 

It is well known that weak corrections can also be significant because
of the presence of possible large Sudakov logarithms. These effects were studied intensively
for several processes like weak boson, jet and top-quark pair production
\cite{Kuhn:2005az,Kuhn:2005gv,Kuhn:2007cv,Moretti:2006nf,Moretti:2006ea,Beenakker:1993yr,Kuhn:2005it,Kuhn:2006vh,Bernreuther:2006vg}. 
Earlier work on Sudakov logarithms in four-fermion processes can be found in
\Refs{Kuhn:1999de,Kuhn:1999nn,Kuhn:2001hz,Feucht:2004rp,Jantzen:2005az,Bec,CiaCom,Fad,DenPoz}. 
A first study for $\bbb$ production can be found in \Ref{Maina:2003is}, where the
partonic sub-processes $gg\ra \bbb$ and $\qqb\ra \bbb$ were
considered. 

In experimental studies, e.g. of top-quark pair production at the Tevatron, often only
one $b$-jet is required. Considering the importance of these background studies 
also single bottom-quark production will be studied in this paper.

The outline of the paper is as follows: In Section~\ref{sec:leadingorder} the leading-order cross
sections for the production of one and two bottom jets are presented. The partonic 
contributions are split into quark-, gluon- and pure bottom-induced
processes and relations based on crossing symmetries are introduced. 
After a discussion at the hadronic level we find, that
leading order processes involving electroweak gauge boson exchange can be neglected. Concerning the 
differential $\pt$ distribution at high energies the effects from pure bottom-induced contributions are 
also negligible. In Section~\ref{sec:NLO} we present the
virtual and real corrections of order $\alpha_s^2\alpha$ for the remaining
processes. The virtual contributions to quark-induced processes contain infrared and
ultraviolet singularities, while the gluon-induced ones are infrared finite. 
Compact analytic expressions for the various
channels are presented. In
Section~\ref{sec:results} we present various consistency tests of our calculation and discuss the numerical results 
at the hadronic level. Our conclusions are given in Section~\ref{sec:conclusion}. 
\section{Bottom-jet production at leading-order}
\label{sec:leadingorder}
\begin{figure}[!htbp]
  \begin{center}
    \leavevmode
    \includegraphics[width=8.2cm]{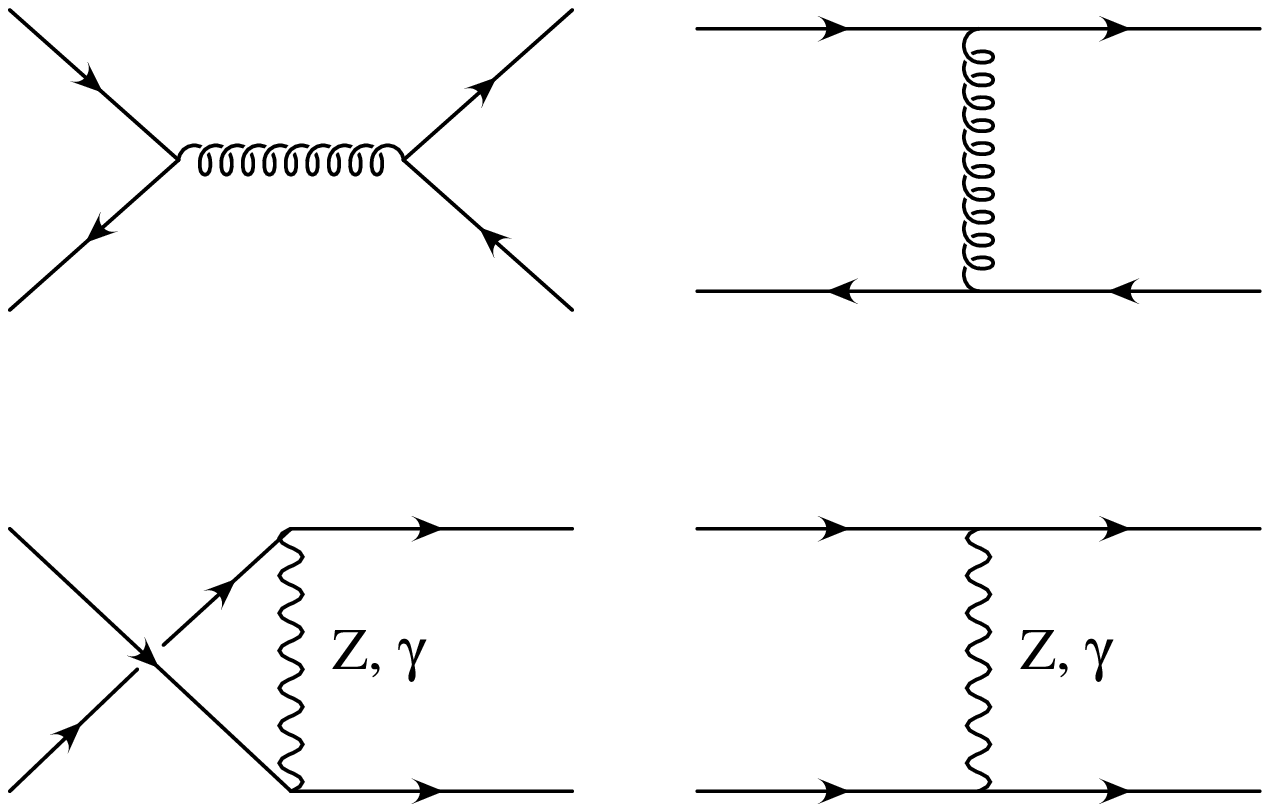}
    \caption{Sample Born diagrams for quark-induced bottom-quark production.}
    \label{fig:quarkborns}
    \vspace*{2cm}
    \includegraphics[width=12.5cm]{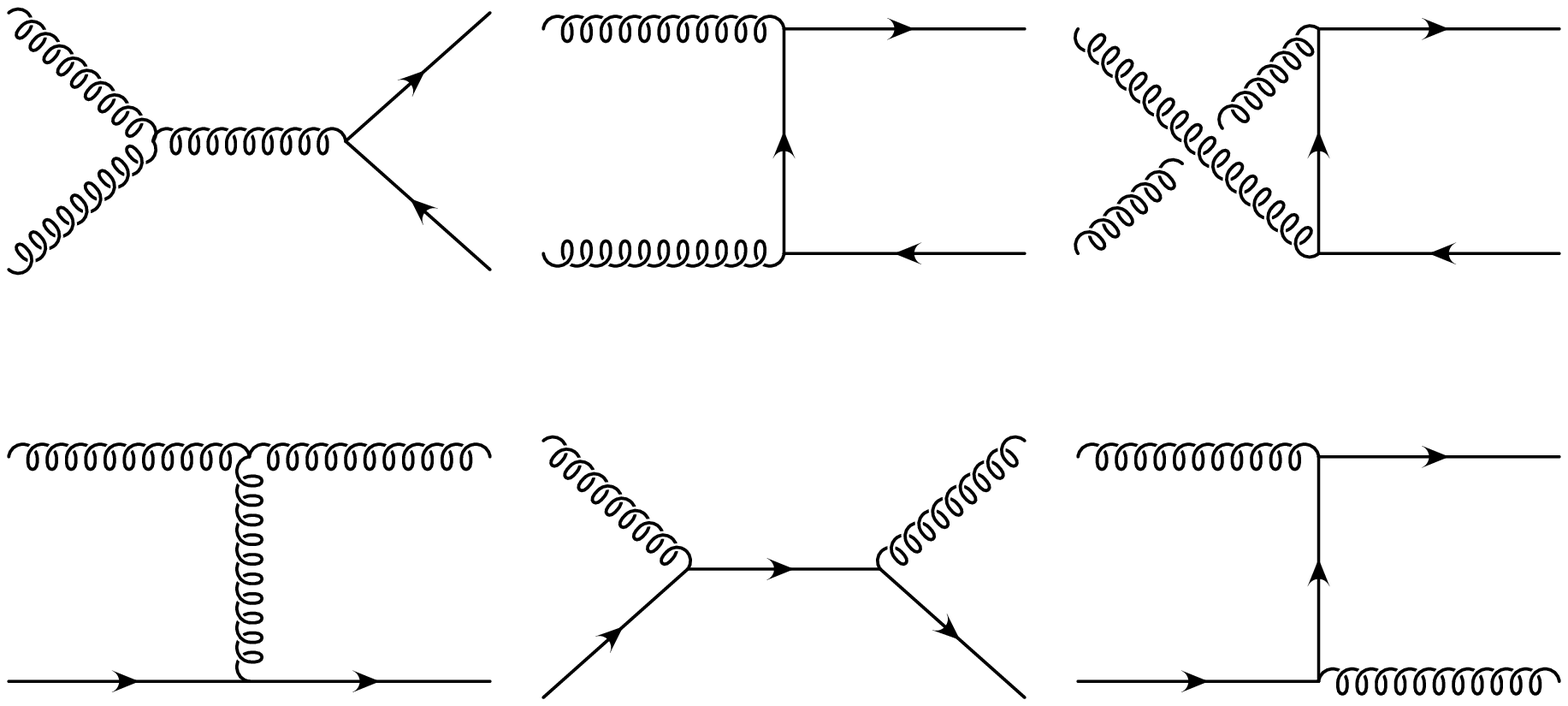}
    \caption{Sample Born diagrams for gluon-induced bottom-quark production.}
    \label{fig:gluonborns}
  \end{center}
\end{figure}

At parton level three types of processes will be distinguished:
Quark-induced processes  (with two quarks in the initial state, one of these being $u,d,c,s$), 
gluon-induced contributions (with one or two gluons in
the initial state) and finally pure bottom-quark induced processes:
\vspace*{-0.2cm}
\begin{eqnarray}
&{\rm quark-induced}:&\qqb\ra\bbb,\:\:\bb q\ra\bb q,\:\: b\qb\ra b\qb,\:\: bq\ra bq,\:\: \bb\qb\ra\bb\qb, \nn\\
&{\rm gluon-induced}:&  gg\ra\bbb,\:\: bg\ra bg,\:\: \bb g\ra\bb g, \nn\\
&{\rm bottom-induced}:& \bbb\ra\bbb,\:\: bb\ra bb,\:\: \bb\bb \ra\bb \bb.
\label{eq:lo-list}
\end{eqnarray}
Processes with initial state photons are neglected. 
Here the light quarks (denoted generically by $q$ and $\qb$) and the bottom-quark
($b$ and $\bb$) are taken as massless. Sample diagrams
are shown in Fig.~\ref{fig:quarkborns} and
Fig.~\ref{fig:gluonborns}. 
The colour and spin averaged squared matrix element $\sumqn|\M^{i,j\ra k,l}|^2 $ can be
written either in terms of the Mandelstam variables $s$, $t$ and $u$
or as a function of $s$ and the cosine of the scattering angle
$\cos\vartheta = z = (u-t)/s$
\begin{equation}
\sumqn|\M^{i,j\ra k,l}|^2(s,t,u) = \sumqn|\M^{i,j\ra k,l}|^2(s,z).
\end{equation}
Various explicitly checked crossing relations will be useful below.
For the quark-induced processes in \Eq{eq:lo-list} we find 
\begin{eqnarray}
\sumqn|\M^{q\bb\ra q\bb}|^2 = \sumqn|\M^{\qb b\ra \qb b}|^2 &=&
\sumqn|\M^{\qqb\ra \bbb}|^2\Big|_{s\lra t}, \nn\\
\sumqn|\M^{qb\ra qb}|^2 = \sumqn|\M^{\qb \bb\ra \qb \bb}|^2 &=&
\sumqn|\M^{\qb b\ra \qb b}|^2\Big|_{s\lra u} \nn\\ 
&=&\sumqn|\M^{\qqb\ra \bbb}|^2\Big|_{s\ra t,\:\: t\ra -s,\:\: u\ra -u}
\label{eq:quark-cross}
\end{eqnarray}
and similarly for gluon-induced and pure bottom-induced
processes 
\begin{equation}
\sumqn|\M^{gb\ra gb}|^2 = \sumqn|\M^{g\bb\ra g\bb}|^2 = -\sumqn|\M^{gg\ra\bbb}|^2\Big|_{s\lra t},
\label{eq:gluon-cross}
\end{equation}
\begin{equation}
\sumqn|\M^{bb\ra bb}|^2 = \sumqn|\M^{\bb\bb\ra \bb\bb}|^2 = \sumqn|\M^{\bbb\ra\bbb}|^2\Big|_{s\lra u}.
\label{eq:bottom-cross}
\end{equation}
The spin and colour averaged differential partonic cross section reads 
\begin{equation}
{\ind\sigma^{i,j\ra k,l}\over \ind z} = {1\over I}{1\over 32\pi s}\sumqn|\M^{i,j\ra k,l}|^2(s,z)
\label{eq:def-cross-section}
\end{equation}
where $I=2$ for identical particles in the final state and $I=1$
otherwise.\\
Before presenting explicit results we complete the notation
and definitions used in this and the following Sections. 
As usual $\alpha_s$ and $\alpha$ stand for
the strong and electromagnetic coupling constant. The $Z$ boson
mass is denoted by $\mz$ and the (axial-)vector coupling is expressed in
terms of the sine (cosine) of the weak mixing-angle $\sw$ ($\cw$), the weak isospin
$T_3^f$ and the electric charge $Q_f$ for a fermion of
flavour $f$ in units of the elementary charge $e$:
\begin{eqnarray}
g_v^f &=& {1\over2\sw\cw}(T_3^f-2\sw^2Q_f),\quad g_a^f = {1\over2\sw\cw}T_3^f.
\label{eq:Z_param}
\end{eqnarray}
The Cabibbo-Kobayashi-Maskawa mixing matrix is set to 1 and the $W$ coupling to
the quarks is
\begin{equation}
\gw = {1\over2\sqrt{2}\sw}.
\label{eq:W_param}
\end{equation}\\
Let us start with the processes listed in \Eq{eq:lo-list}, which involve only quarks in the
initial state. 
These can be mediated by gluon, $Z$ boson  
and photon exchange. Consequently we seperate pure QCD $\O(\alpha_s^2)$, 
electroweak $\O(\alpha^2)$, and mixed QCD-electro\-weak $\O(\alpha_s\alpha)$
contributions. In the case of one or two gluons in the initial-state an interaction via electroweak
gauge bosons is not possible at leading order and only QCD processes contribute.
We start with the QCD-contribution for quark--antiquark
annihilation into a $\bbb$-pair
\begin{equation}
\sumqn|\M_{\alpha_s^2}^{\qqb\ra\bbb}|^2\Big(s,z(s,t,u)\Big) = 4\*\alpha_s^2\*\pi^2{N^2-1\over N^2}\*(1+z^2),
\label{eq:Mlo-qqb-bbb}
\end{equation}
with $N$ being the number of colors. The purely electroweak contributions are given by
\begin{eqnarray}
&&\sumqn|\M_{\alpha^2}^{\qqb\ra\bbb}|^2\Big(s,z(s,t,u)\Big) =16\*\alpha^2\*\pi^2\*\Big[\Qb^2\*\Qq^2\*(1+z^2) \nn\\
&&\quad\quad+\quad{s^2\over(s-\mz^2)^2}\*\Big((\gvq^2+\gaq^2)\*(\gvb^2+\gab^2)\*(1+z^2)-8\*\gvq\*\gvb\*\gaq\*\gab\*z\Big)\nn\\
&&\quad\quad+\quad2\*{s\over s-\mz^2}\*\Qq\*\Qb\*\Big(\gvq\*\gvb\*(1+z^2)-2\*\gaq\gab\*z\Big)\Big].
\label{eq:Mlo-weak-qqb-bbb}
\end{eqnarray}
Because of colour conservation the interference between the QCD and electroweak
matrix elements vanish. The remaining quark-induced contributions can be obtained
using the crossing relations in \Eq{eq:quark-cross}. \\
The gluon-induced processes in \Eq{eq:lo-list} proceed through QCD amplitudes only and the corresponding crossing relations are defined in
\Eq{eq:gluon-cross}. For the gluon fusion channel the squared
matrix element is 
\begin{equation}
\sumqn|\M_{\alpha_s^2}^{gg\ra\bbb}|^2\Big(s,z(s,t,u)\Big) = 8\alpha_s^2\*\pi^2\*{1\over N(N^2-1)}\*(N^2-2+N^2z^2){1+z^2\over1-z^2}.
\label{eq:Mlo-gg-bbb}
\end{equation}
\\
Finally we consider the pure bottom-quark induced contributions arising from pure QCD, mixed
QCD-electro\-weak and electroweak matrix elements. Below we list the results for
$\bbb\ra\bbb$ scattering, those for $bb\ra bb$ and $\bb\bb\ra\bb\bb$ scattering
can be deduced from \Eq{eq:bottom-cross}.
\begin{eqnarray}
\sumqn|\M_{\alpha_s^2}^{\bbb\ra\bbb}|^2\Big(s,z(s,t,u)\Big) &=& 4\*\alpha_s^2\*\pi^2\*{N^2-1\over N^3}\nn\\
&\times&{2\*(1-z)\*(1-z^2)+N\*(11-2\*z+4\*z^2+2\*z^3+z^4)\over(1+z)^2},\nn\\
\label{eq:Mlo-bbb-bbb}
\end{eqnarray}
\begin{eqnarray}
\sumqn|\M_{\alpha_s\alpha}^{\bbb\ra\bbb}|^2\Big(s,z(s,t,u)\Big) &=&
-16\*\alpha\*\alpha_s\*\pi^2\*{N^2-1\over  N^2}\*\Big[2\*\Qb^2\*{(1-z)^2\over1+z}\nn\\
&+&(\gvb^2+\gab^2)\*{s\over
  s-\mz^2}\*{(1-z)^2\over1+z}\*{2\*s\*(1+z)+\mz^2\*(1-z)\over s\*(1+z)+2\*\mz^2} \Big],\nn\\
\end{eqnarray}
\begin{eqnarray}
\sumqn|\M_{\alpha^2}^{\bbb\ra\bbb}|^2\Big(s,z(s,t,u)\Big) &=&
16\*\alpha^2\*\pi^2\nn\\
&\times&\Bigg[\Qb^4\*{N\*(11-2\*z+4\*z^2+2\*z^3+z^4)-2\*(1-z)\*(1-z^2)\over N\*(1+z)^2}\nn\\
&+&{s^2\over N\*(s-\mz^2)^2\*(s\*(1+z)+2\*\mz^2)^2}\nn\\
&\times&\Big[(\gvb^4+\gab^4)\*\Big(2\*\mz^4\*(2\*(1-z)^2+N\*(7-2\*z+3\*z^2))\nn\\
&-&2\*s\*\mz^2\*(1-z)\*((1-z)^2+2\*N\*(4+z+z^2))\nn\\
&-&s^2\*(2\*(1+z)\*(1-z)^2-N\*(11-2\*z+4\*z^2+2\*z^3+z^4))\Big)\nn\\
&+&\gab^2\*\gvb^2\*\Big(4\*\mz^4\*(6\*(1-z)^2+N\*(1-14\*z+5\*z^2))\nn\\
&-&4\*s\*\mz^2\*(1-z)\*(3\*(1-z)^2-2\*N\*(2+5\*z-z^2))\nn\\
&-&2\*s^2\*(6\*(1+z)\*(1-z)^2+N\*(1+14\*z+2\*z^3-z^4))\Big)\Big]\nn\\
&-&2\*\Qb^2\*{s\over N\*(1+z)\*(s-\mz^2)\*(s\*(1+z)+2\mz^2)}\nn\\
&\times&\Big[\gvb^2\*\Big(\mz^2\*\*(1-z)\*((1-z)^2+2\*N\*(4+z+z^2))\nn\\
&+&s\*(2\*(1+z)\*(1-z)^2-N\*(11-2\*z+4\*z^2+2\*z^3+z^4))\Big) \nn\\
&+&\gab^2\*\Big(\mz^2\*(1-z)\*((1-z)^2-6\*N\*(1+z))\nn\\
&+&2\*s\*(1+z)\*((1-z)^2+N\*(3+z^2))\Big)\Big]\Bigg].
\end{eqnarray}
The term proportional to $\alpha\as$ originates from the interference between $s$ and $t$ exchange amplitudes and is present for
the process $\bbb\ra\bbb$ and its crossed versions only. The purely electroweak contributions proportional $\alpha^2$ are found 
to be negligible.
\\
With these ingredients we calculate the hadronic transverse momentum ($\pt$) distributions
for single and double $b$-tag events and investigate the relative importance of the different processes 
described above. As a consequence of the massless approximation, most partonic
contributions are ill-defined in the limit $z\to \pm 1$, where initial and final
state parton become collinear. Since $b$-jets close to the beam 
pipe escape detection, we require a minimal transverse momentum ($\ptcut$) of 
50 GeV. The leading order cross section is obtained from 
\pagebreak
\begin{eqnarray}
\ind\sigma_{H_1,H_2\ra X} &=& \sum_{i,j}\int_0^1 \ind x_1\int_0^1 \ind x_2
%\Theta(x_1x_2-\tau_{i,j})
f_{i/H_1}(x_1)f_{j/H_2}(x_2) \nn\\
&\times& 
\ind\hat{\sigma}_{i,j\ra X}(x_1P_1,x_2P_2)\:\:\:\Theta(\pt > \ptcut),
\label{eq:Hadron-WQ}
\end{eqnarray}
where the factorization scale dependence is suppressed and $x_1$ and $x_2$ are the
partonic momentum fractions. The parton
distribution functions (PDF's) for parton $i$ in hadron $H$ are denoted by $f_{i/H}$. The sum runs over all possible
parton configurations $(i,j)$ in the initial state. 

For the study of $b$-jets we will distinguish 
between single $b$-tag and double $b$-tag events, and evaluate the $\pt$ distributions with the following
input parameters
\begin{equation}
\alpha_s = 0.1, \quad \alpha = {1\over 128},\quad \mz = 91.1876~{\rm GeV}, \quad \mw = 80.425~{\rm GeV}.
\end{equation}
For $\sw$ we use the on-shell relation. Moreover we used the
PDF's from CTEQ6L with the factorisation scale set to $\muF = 2\ptcut$. 
\begin{figure}[!htb]
  \begin{center}
    \leavevmode
    \hspace*{-0.3cm}
    \begin{minipage}{15cm}
    \includegraphics[width=0.49\textwidth]{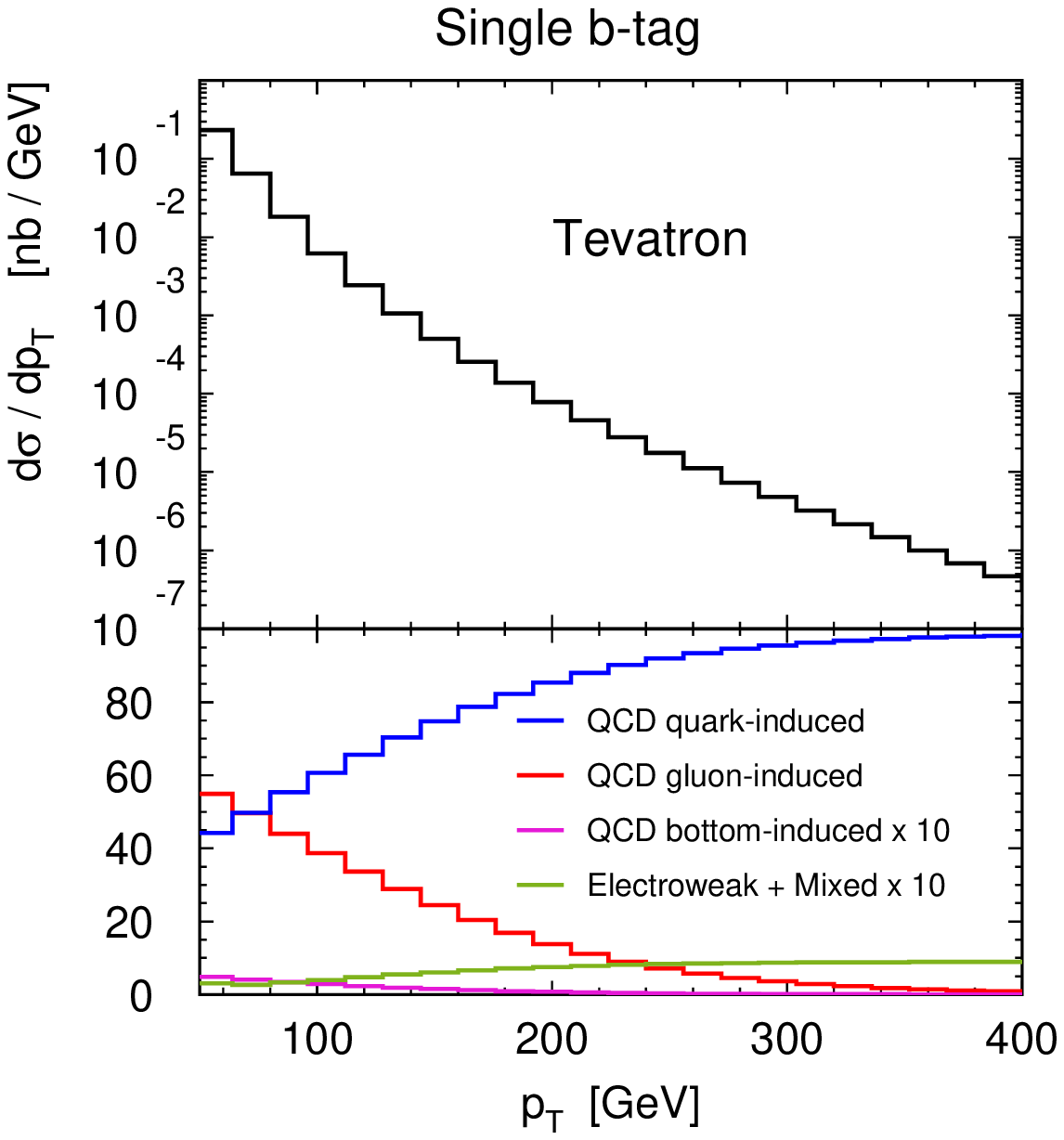}
    \includegraphics[width=0.49\textwidth]{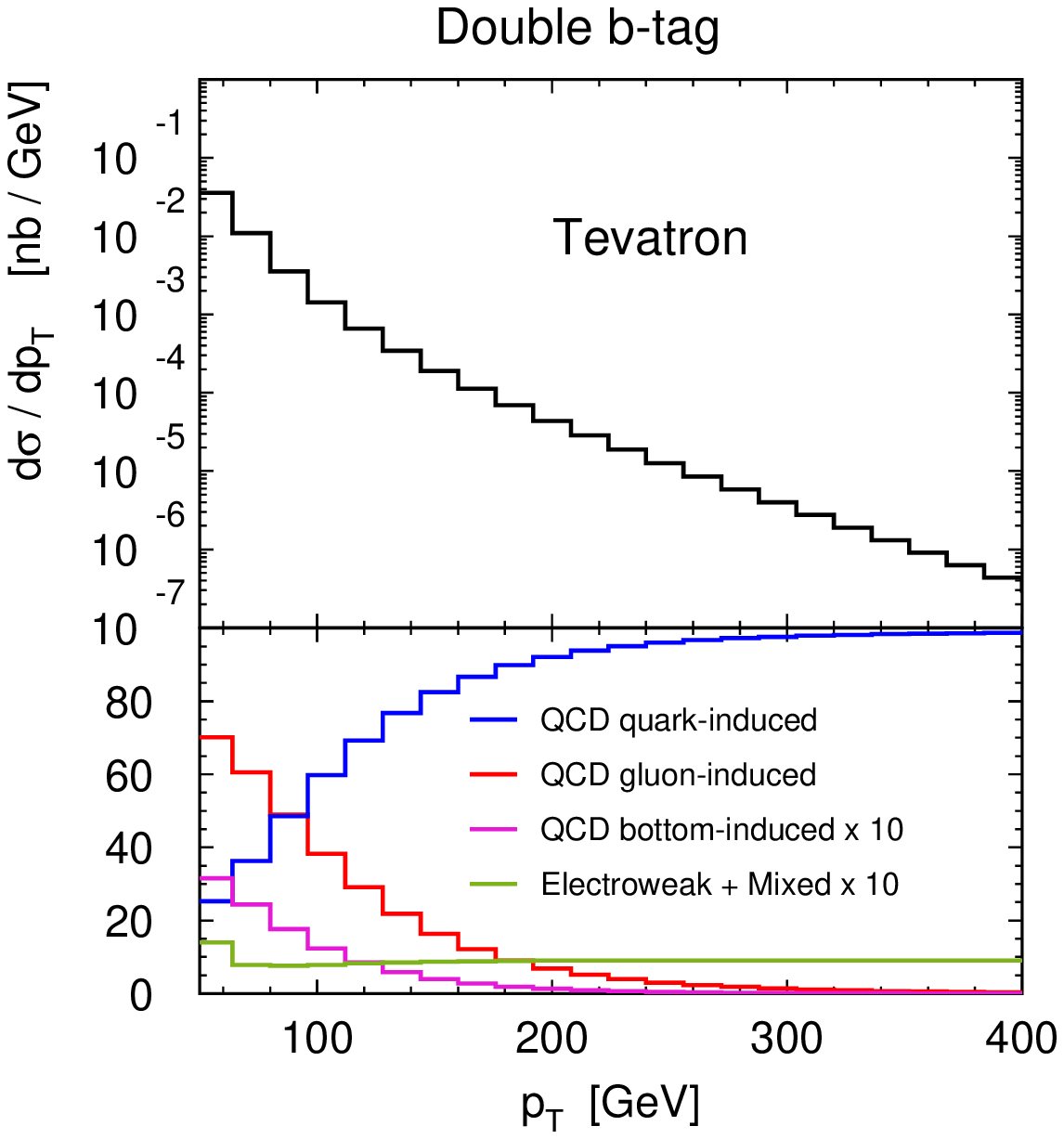}
    \end{minipage}
    \rput(-1.19,6.98){\large (a)}
    \rput(6.27,6.98){\large (b)}
     \caption{Differential cross section as a function of
       $\pt$ for single $b$-tag events (upper left figure) and double $b$-tag events (upper right figure) 
       at the Tevatron ($\sqrt{s} = 1.96 $ TeV) and the relative composition normalized to the Born cross section (lower
       figures)}
     \label{fig:pt-bjet-tevatron}
  \end{center}
\end{figure}
\begin{figure}[!htb]
  \begin{center}
    \leavevmode
    \hspace*{-0.4cm}
    \begin{minipage}{15cm}
    \includegraphics[width=0.49\textwidth]{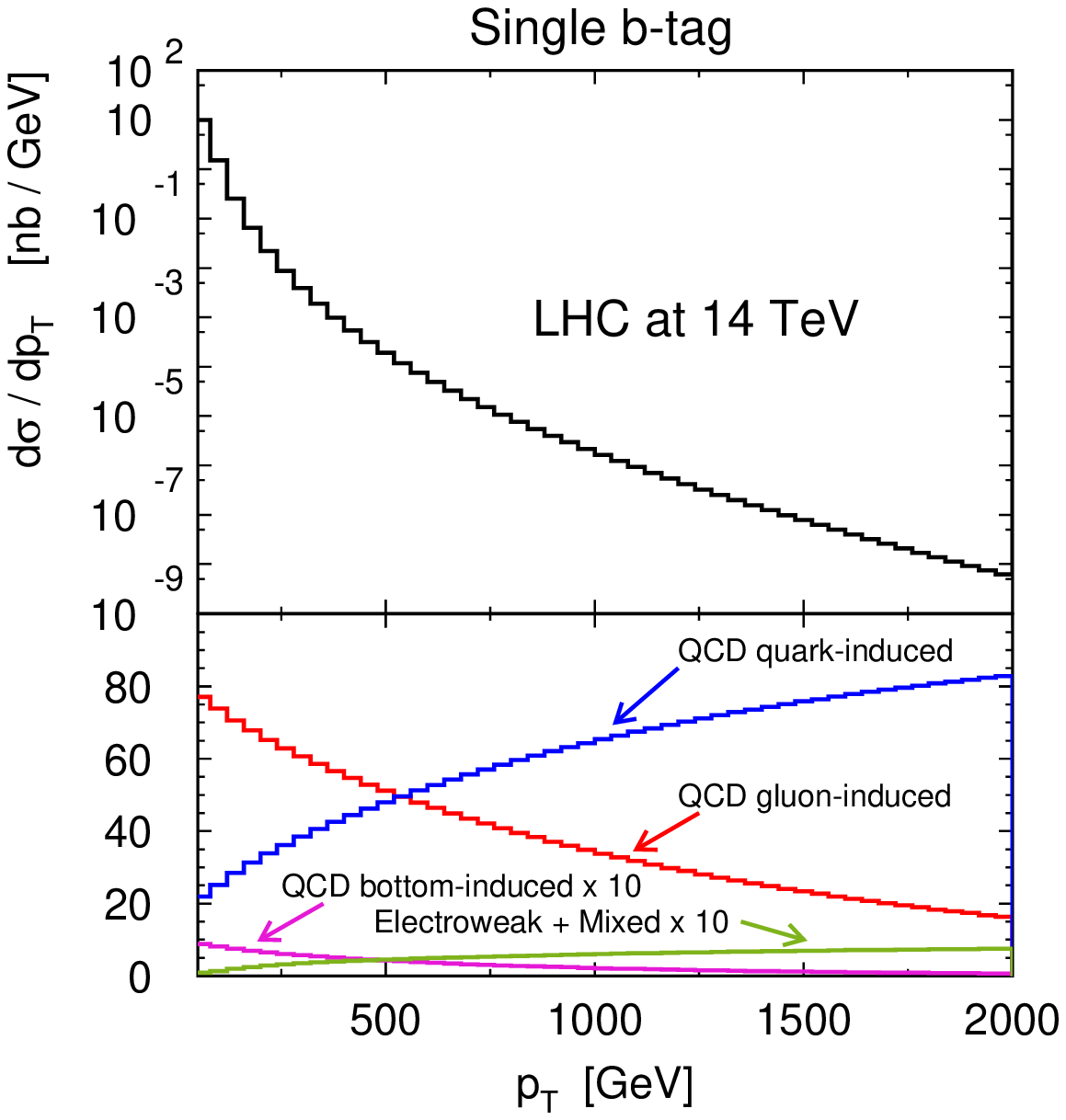}
    \includegraphics[width=0.49\textwidth]{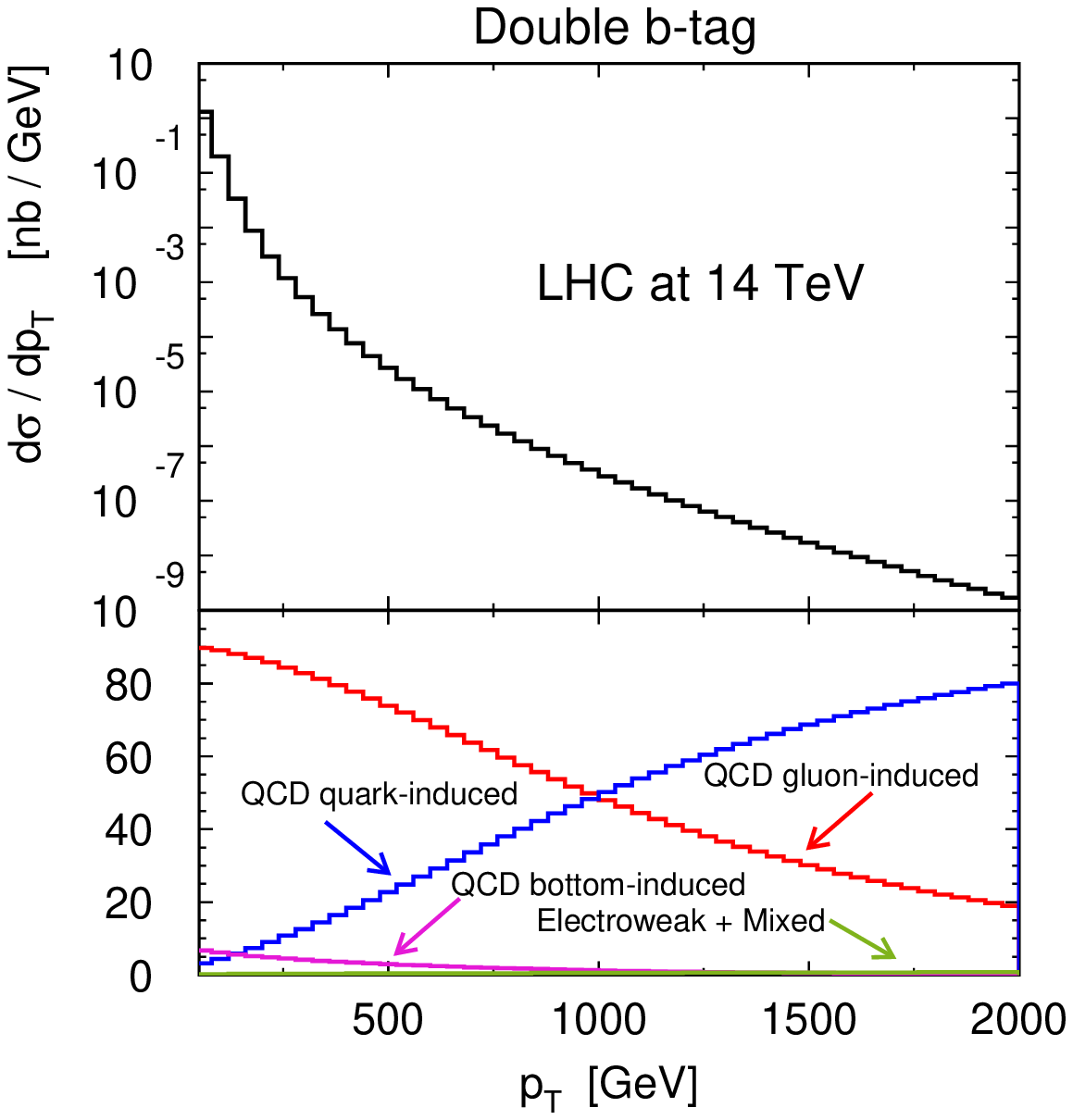}
    \end{minipage}
    \rput(-1.2,7.03){\large (a)}
    \rput(6.28,7.03){\large (b)}
    \caption{Differential cross section as a function of
       $\pt$ for single $b$-tag events (upper left figure), double $b$-tag events (upper right figure) 
       at the LHC ($\sqrt{s} = 14 $ TeV) and the relative composition normalized to the Born cross section (lower
       figures)}
    \label{fig:pt-bjet-lhc}
  \end{center}
\end{figure}
\\
Let us first discuss the results for the Tevatron ($\sqrt{s} = $ 1.96 TeV).
The composition of the corresponding leading-order differential cross 
section for single $b$-tag events is shown in Fig.~\ref{fig:pt-bjet-tevatron}(a). In the upper plot 
the cross section is depicted as a function of the transverse momentum $\pt$, 
its relative composition in percent is shown below. For $\pt$-values up to 200 GeV the 
gluon- and quark-induced QCD processes are of the same order, while  
for higher energies the quark-induced $b$-jet production is dominating the $\pt$-distribution.
Contributions from gluon-induced processes drop fast with increasing $\pt$ and contribute only a few percent to 
the differential cross section for $\pt > 300$ GeV. The relative contributions from bottom-induced processes and processes with electroweak boson exchange are only of 
the order of few permille and therefore negligible for the $\pt$-values considered. \\
For $\pt < 100$ GeV and in the double $b$-tag case the $\pt$-distribution (Fig.~\ref{fig:pt-bjet-tevatron}(b))
is dominated by the gluon fusion channel. For higher $\pt$-values
QCD quark--antiquark annihilation takes over and for transverse momenta larger than 200 GeV it dominates completely. 
The relative contributions from processes involving electroweak bosons are of the order of $1\%$
for $\pt< 80$ GeV and, for $\pt > 260$ GeV are of the same order (few permille) as the gluon-fusion process. Bottom-induced contributions yield a few percent to 
the differential cross section for $\pt < 100$ GeV and become insignificant for higher transverse momenta.\\
For the LHC the results are significantly different. 
The LO $\pt$-distribution for single $b$-tag events is presented in Fig.~\ref{fig:pt-bjet-lhc}(a).
The gluon-induced processes dominate in the low energy regime 
($\pt < 500$ GeV). For $\pt$ larger than 500 GeV the 
quark-induced processes take over and finally dominate the
distribution. This "cross over" of gluon- and quark-induced contributions is a consequence of 
the different behaviour of LHC quark and gluon luminosities.  
The relative contributions from the exchange of electroweak bosons and 
bottom-induced processes are always below $1\%$ and therefore negligible.
A similiar picture is observed for the differential cross section for
double $b$-tag events at the LHC (Fig.~\ref{fig:pt-bjet-lhc}(b)). Here the "cross over" of quark- and 
gluon-induced contributions is around $\pt = 1$ TeV. 
For low $\pt$ the pure bottom-induced processes are as important as quark-induced contributions. 
This seems surprising, because the parton luminosities of bottom-quarks
in a proton should be highly suppressed compared
to the light flavours. There are two reasons for the
relatively large cross section of the purely bottom-induced processes. 
First, the partonic cross sections of $bb$ and 
$\bbb$ scattering are strongly enhanced for large $z$ values. In this region the parton 
processes with bottom-quarks in the initial state 
can be several orders of magnitude larger than the quark--antiquark-induced process. Second, the
bottom-quark PDF is essentially obtained by multiplying the gluon distribution in the
proton with the splitting function of a gluon into a bottom-quark pair. 
Because of the high gluon luminosities at low energies, 
the bottom-quark PDF becomes of the order of a few percent relative to the PDF's of the light
flavours. This, together with the large partonic contributions is responsible for
the relatively large bottom-induced differential cross section. The argumentation implies 
that the main contribution from $bb$, $\bb\bb$ and $\bbb$ scattering comes from the low $\pt$ region,
while for high $\pt$ values these effects are small. It might be interesting to study whether the
$b$-PDF could be further constrained using $\bbb$ production at low $\pt$.
\\

As shown above, the leading order contributions from electroweak
gauge boson exchange are always negligible for the study of $b$-jet production. This is also true in the context of
NLO corrections with an expected size of serveral percent relative to
the leading order distributions. Moreover, we have shown that the QCD contributions from processes with two bottom-quarks
in the initial state are unimportant for the study of $\pt$-distributions at large transverse momentum. 
In particular with regard to the Sudakov logarithms becoming important at high energies this approximation is 
justified. Consequently the weak $\order{\alpha}$ corrections to $bb\ra bb$, 
$\bb\bb\ra\bb\bb$ and $\bbb\ra\bbb$ will not be discussed in 
this article.
\newpage
\section{Weak corrections to bottom-quark production}
\label{sec:NLO}
In this Section we calculate the weak corrections of order
$\alpha_s^2\alpha$ to the following partonic processes
\vspace*{-0.3cm}
\begin{eqnarray}
&& \:\: \qqb\ra\bbb,\:\:\bb q\ra\bb q,\:\: b\qb\ra b\qb,\:\: bq\ra bq,\:\: \bb\qb\ra\bb\qb, \nn\\
&& \:\: gg\ra\bbb,\:\: bg\ra bg,\:\: \bb g\ra\bb g, \nn
\label{eq:nlo-list}
\end{eqnarray}
neglecting photonic corrections. We subdivide the $\order{\alpha}$ corrections in contributions from
quark-induced processes (Section~\ref{sec:q-NLO}) and gluon-induced processes
 (Sec\-tion \ref{sec:g-NLO}). Before presenting analytic results for the weak corrections, let us
add some technical remarks. For the calculation of the next-to-leading order
weak corrections the `t Hooft-Feynman gauge with gauge parameters set to
1 is used. The longitudinal degrees of freedom of the massive gauge bosons $Z$ and $W$
are thus represented by the goldstone fields $\chi$ and $\phi$. As mentioned
in Section~\ref{sec:leadingorder}, all incoming and outgoing
partons are massless and consequently there are no
contributions from the Goldstone boson $\chi$ and the Higgs boson. Ghost fields do
not contribute at the order under consideration. For the analytic reduction of the tensor
integrals to scalar integrals we used the Passarino-Veltman reduction
scheme \cite{Passarino:1978jh}. The scalar integrals were calculated either analytically with standard
techniques or evaluated numerically using the FF-library
\cite{vanOldenborgh:1990yc}. The convention for the scalar
integrals in this article is:
\begin{equation}
  {\rm X}_0
  = {1\over i\pi^2} \int \ind^d\ell
  {(2\pi\mu)^{2\e}\over (\ell^2-m_1^2+i\epsilon)\cdots}.
\end{equation}
The bare Lagrangian $\cal L$ is
rewritten in terms of renormalised fields and couplings as follows
\begin{eqnarray}
  {\cal L}(\Psi_0,A_0, m_0,g_0) &=&
  {\cal L}(Z^{1/2}_\Psi\Psi_R,Z^{1/2}_A A_R, Z_m m_R,Z_g g_R)\nn\\
  &\equiv&  {\cal L}(\Psi_R,A_R, m_R,g_R)
  + {\cal L}_{ct}(\Psi_R,A_R, m_R,g_R)
  \label{eq:RenormalizedPerturbationTheory}
\end{eqnarray}
(see e.g. \Ref{Denner:1991kt}). 
For the present calculation only wave function renormalization is
needed and no mass or coupling renormalization has to be performed. The  
partonic NLO corrections are, furthermore, independent of the factorization scale $\mu_F$. 
The wave function renormalization is performed in the on-shell scheme
\begin{eqnarray}
  \Psi^{\rm R,L}_{0} &=&  \L Z^{\rm R,L} \R ^{1/2} \Psi^{\rm R,L}
  =\L 1 + {1\over 2} \delta Z^{\rm R,L} \R  \Psi^{\rm R,L},
\end{eqnarray}
where the renomalization constants are given in terms of
self-energy corrections $\Sigma$ and their derivatives:
\begin{eqnarray}
  \delta Z_{\rm V} &=& {1\over 2}(\delta Z^{\rm L}+\delta Z^{\rm R}), \nn\\
  \delta Z_{\rm A} &=& {1\over 2}(\delta Z^{\rm L}-\delta Z^{\rm R}). 
\end{eqnarray}
Here only $\delta Z_{\rm V}$ is required. 
For bottom-quarks it is given by \cite{Denner:1991kt}:
\begin{eqnarray}
\delta Z_{\rm V}^b &=& {\alpha\over4\*\pi}\*\Bigg[(\gvb^2+\gab^2)\*
\left({3\over2}-{1\over\mz^2}\*{\rm A}_0(\mz^2)\right) \nn\\
&+&2\*\gw^2\*\Bigg(1+{1\over\mt^2-\mw^2}\*\left({\rm A}_0(\mw^2)-{\rm A}_0(\mt^2)\right)
\nn\\
&-&\*(\mt^2-\mw^2)\*{\partial\over \partial p^2}\*{\rm B}_0(p^2,\mt^2,\mw^2)\Big|_{p^2=0}\Bigg)\nn\\
&+&\gw^2\*{\mt^2\over\mw^2}\*\Bigg({1\over{\mt^2-\mw^2}}\*\left({\rm A}_0(\mw^2)-{\rm A}_0(\mt^2)\right)\nn\\
&-&\*(\mt^2-\mw^2)\*{\partial\over \partial p^2}\*{\rm B}_0(p^2,\mt^2,\mw^2)\Big|_{p^2=0}\Bigg)\Bigg]
\label{eq:bottom_count}
\end{eqnarray}
corresponding to the massless limit of the formulae given in \Ref{Kuhn:2006vh}
after the exchange of bottom and top mass. For light quark flavours the renormalisation constant reads:
\begin{eqnarray}
\delta Z_{\rm V}^q &=& {\alpha\over4\*\pi}\*\Bigg[(\gvq^2+\gaq^2)\*
\left({3\over2}-{1\over\mz^2}\*{\rm A}_0(\mz^2)\right) \nn\\
&+&2\*\gw^2\*\Big({3\over2}-{1\over\mw^2}\*\*{\rm A}_0(\mw^2)\Big)\Bigg].
\label{eq:quark_count}
\end{eqnarray}
\\
In the following, most results will be given using the explicit formulae of
the scalar integrals. To evaluate the NLO corrections to 
the differential cross section only the real part of the virtual corrections contribute.  
Therefore we will skip the parameter for analytic
continuation $\epsilon$ wherever possible. Hence the relations from 
crossing symmetries lead for example to the following replacements ($s>0$, $t<0$, $M^2>0$) 
\begin{eqnarray*}
&&\ln\left({s\over M^2}\right) \stackrel{s\lra t}
{\longrightarrow} \ln\left({|t|\over  M^2}\right),\quad \ln^2\left({s\over M^2}\right) \stackrel{s\lra t}{\longrightarrow}
 \ln^2\left({|t|\over
  M^2}\right) -\pi^2\nn.\\
\end{eqnarray*}
\subsection{Weak corrections to quark-induced processes}
\label{sec:q-NLO} 
\begin{figure}[!htbp]
  \begin{center}
    \leavevmode
    \includegraphics[width=9.2cm]{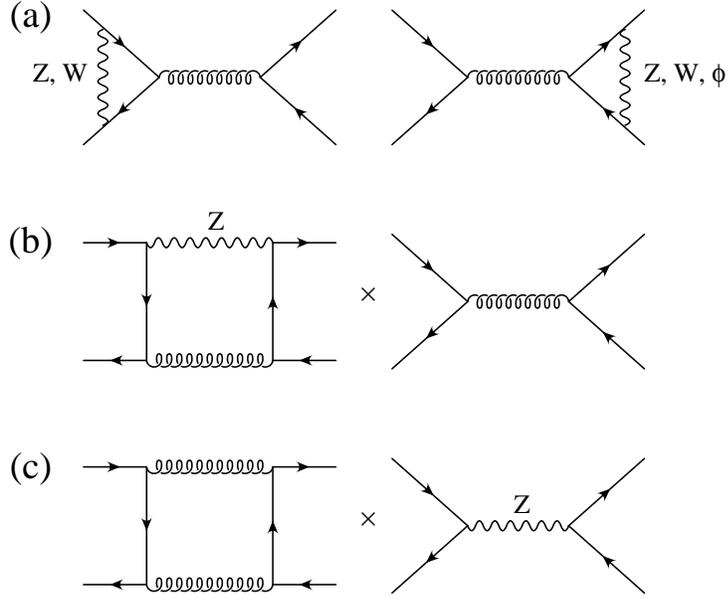}
    \caption{Sample diagrams for the quark--antiquark annihilation process at $\order{\alpha_s^2\alpha}$.}
    \label{fig:qqb-nlo}
  \end{center}
\rput(2.6,9.5){{\large (a)}}
\rput(2.6,6.5){{\large (b)}}
\rput(2.6,3.5){{\large (c)}}
\end{figure}

We start with the $\order{\alpha}$ corrections to quark-induced processes, 
consisting of vertex-, box- and real-contributions. In contrast to the gluon-induced reaction
where electroweak corrections to the QCD Born amplitude can be clearly identified, the corresponding classification 
for $\qqb\ra\bbb$ more precisely of all contributions of $\order{\aas}$ is more involved. In this case QCD box amplitudes 
$\order{\as^2}$ may interfere with the weak amplitude of $\order{\alpha}$ and similarly, mixed box amplitudes of $\order{\as\alpha}$
may interfere with the QCD Born amplitude of $\order{\as}$. Infrared singularities are cancelled by contributions from interference 
between initial state radiation and final state radiation. A closely related discussion can be found in \Ref{Kuhn:2005it} 
and for QED corrections to neutral current corrections in \Refs{Kuhn:1987nh,Jadach:1987ws}.
Two types of box-diagrams can be distinguished:
\begin{enumerate}
\item The (box-type) weak correction to the QCD Born amplitude, 
  interfering with the QCD Born amplitude.
\item The QCD box diagram interfering with the weak Born
  amplitude.
\end{enumerate}
Sample diagrams are shown in Fig.~\ref{fig:qqb-nlo} b), c). The box-diagrams are UV finite. 
Their IR singularities cancel against those from real emission, specifically, 
from interference terms between initial and final state radiation.
The vertex-corrections are infrared (IR) finite
and their UV divergencies cancel against the wave function renormalization described above. 
All contributions are free from initial state mass singularities. 
To handle the infrared singularities we use the dipole subtraction method \cite{Catani:1996vz}. 
In the notation of \Ref{Catani:1996vz} the NLO box and real corrections can be written as
\begin{eqnarray}
\delta\sigma^{{\rm NLO} \:\:\order{\alpha_s^2\alpha}}\Big|_{\rm\Box + Real} &=& 
\Bigg[\ind\sigma^{\rm\Box} + \left( {\bf I}\:\:\: \otimes \:\:\: \ind\sigma^{\BB}\right)\Bigg]\nn\\
&+& \Bigg[\ind\sigma^{\rm Real} - \left(\sum_{\rm Dipoles} \ind V_{\rm Dipoles}\:\:\: \otimes \:\:\: \ind\sigma^{\BB}\right)\Bigg]\nn\\
&+&\Bigg[ \int_0^1 \ind x \Big({\bf K}(x)+{\bf P}(x)\Big)\:\:\:\otimes\:\:\: \ind\sigma^{\BB}(x)\Bigg].
\label{eq:qqb-bbb-nlo}
\end{eqnarray}
Here ${\rm d}\sigma^{\rm \Box}$ consists of the virtual box
corrections. Their IR-singularities will be cancelled by the contribution involving the ${\bf I}$-operator. Note that the differential 
cross section $\ind\sigma^{\BB}$ is {\it not} the QCD leading order
contribution. In order to respect the mixed QCD-weak structure of the box-diagrams and of the real corrections, $\ind\sigma^{\BB}$
is given by the interference between the QCD amplitude times the weak
amplitude, where the colour correlations of the dipole formalism yield a
non-vanishing colour factor. We use the symbol `$\otimes$' in Eq.~(\ref{eq:qqb-bbb-nlo}) to denote color as well as spin correlations.
 The ${\bf K}$ and ${\bf P}$ operators are finite remainders originating
from dipoles, for details we refer to \Ref{Catani:1996vz}.\\
For the virtual and real corrections we present analytical results for the quark--antiquark 
annihilation process, while for the remaining quark-induced processes we refer to the crossing relations in Eq.~(\ref{eq:quark-cross}). The results for 
\begin{equation}
({\bf I}+{\bf K}+{\bf P})\otimes \ind\sigma^{\BB}
\end{equation}
will be given explicitly for each process. 
The virtual contributions $\ind\sigma^{\rm Virtual}$ to the partonic differential cross section is given with \Eq{eq:def-cross-section}
\begin{eqnarray}
{\ind\sigma^{\order{\alpha_s^2\alpha} \qqb\ra\bbb}_{\rm Virtual}\over \ind z} &=& {1\over32\pi s}\Bigg[
\sum_{i=Z,W}  \sumqn\left|\M^{V_i(\qqb\ra\bbb)}_{{\rm initial}}\right|^2+ 
\sum_{i=Z,W,\phi} \sumqn\left|\M^{V_i(\qqb\ra\bbb)}_{{\rm final}}\right|^2 \nn\\
&+&  \sumqn\left|\M^{\Box{(\qqb\ra\bbb)}}_{\rm QCD}\right|^2 
+ \sumqn\left|\M^{\Box{(\qqb\ra\bbb)}}_{\rm EW}\right|^2\Bigg].
\end{eqnarray}
The initial-state vertex corrections for $\qqb\ra\bbb$ read 
\begin{eqnarray}
 \sumqn\left|\M^{V(\qqb\ra\bbb)}_{{\rm initial}}\right|^2 &=&
-\sumqn\left|\M^{\qqb\ra\bbb}_{{\alpha_s^2}}\right|^2\nn\\
&\times&{\alpha\over 2\pi}\Bigg[{1\over2}(\gvq^2+\gaq^2) \* f_1\left({\mz^2\over s}\right)+\gw^2\*
f_1\left({\mw^2\over s}\right)\Bigg] 
\end{eqnarray}
with 
\begin{equation}
  f_1(x) = 1+2\*\Big[\big(1+\ln(x)\big)
  \*\big(2\*x+3\big)-2\*\big(1+x\big)^2
  \Big(\Li2{1+{1\over x}}-{\pi^2\over 6}\Big)\Big] \nn
\end{equation}
and the squared leading order matrix element is given in
\Eq{eq:Mlo-qqb-bbb}. The final state vertex corrections read
\begin{equation}
 \sumqn\left|\M^{V_Z(\qqb\ra\bbb)}_{{\rm final}}\right|^2 = -\sumqn\left|\M^{\qqb\ra\bbb}_{{
 \alpha_s^2}}\right|^2{\alpha\over 4\pi}\*(\gvb^2+\gab^2) \* f_1
\left({\mz^2\over s}\right),
\end{equation}
\begin{eqnarray}
\sumqn\left|\M^{V_W(\qqb\ra\bbb)}_{{\rm final}}\right|^2   &=& \sumqn\left|\M^{\qqb\ra\bbb}_{{\alpha_s^2}}\right|^2\*{\alpha\over\pi}\*\gw^2\nn\\
&\times&\*\Bigg\{-1\nn+{2\*\mt^2-2\*\mw^2-3\*s\over s}\nn\\
&\times&\bigg({1\over\mt^2-\mw^2}\*\left(\Amw-\Amt\right)+\*\Bosmtmt\bigg)\nn\\
&-&{2\over s}\*\bigg((\mt^2-\mw^2)^2+s\*(s-\mt^2+2\*\mw^2)\bigg)\*\Cosmtmwmt\nn\\
&-&(\mt^2-\mw^2)\*\DBoomtmw\Bigg\},
\end{eqnarray}
\begin{eqnarray}
\sumqn\left|\M^{V_{\phi}(\qqb\ra\bbb)}_{{\rm final}
 }\right|^2 &=& \sumqn\left|\M^{\qqb\ra\bbb}_{{\alpha_s^2}}\right|^2\*{\alpha\over\pi}\*\gw^2\*{\mt^2\over2\mw^2}\nn\\
&\times&\*\Bigg\{-1\nn+{2\*\mt^2-2\*\mw^2+s\over s}\nn\\
&\times&\bigg({1\over\mt^2-\mw^2}\*\left(\Amw-\Amt\right)+\*\Bosmtmt\bigg)\nn\\
&-&{2\over s}\*\bigg((\mt^2-\mw^2)^2+s\*\mt^2\bigg)\*\Cosmtmwmt\nn\\
&-&(\mt^2-\mw^2)\*\DBoomtmw\Bigg\},
\end{eqnarray}
where the scalar integrals are defined in the Appendix \ref{sec:integrals}. For the box-diagrams we find
\begin{eqnarray}
\sumqn\left|\M^{\Box{(\qqb\ra\bbb)}}_{\rm QCD}\right|^2 &=& 4\*\pi\*\alpha\*\alpha_s^2\*{s\over s-\mz^2}\*{N^2-1\over N^2}\*\nn\\
&\times&\Bigg\{\gvq\*\gvb\*
\Big[\ln\left({t\over u}\right)-z\*\ln\left({t\*u\over s^2}\right)\nn\\
&+&(1+z+z^2)\*\ln^2\left(-{t\over s}\right)-(1-z+z^2)\*\ln^2\left(-{u\over s}\right)\Big]\nn\\
&+&\gaq\*\gab\*\Big[z\*\ln\left({u\over t}\right)+\ln\left({t\*u\over s^2}\right)\nn\\
&-&z\*\ln^2\left(-{t\over s}\right)+z\*\ln^2\left(-{u\over s}\right)\Big]\Bigg\}\nn\\
&+&\*\pi\*\alpha\*\alpha_s^2\*{N^2-1\over  N^2}\*\BB^{\qqb\ra\bbb}\*\Bigg\{(4\pi)^{\veps}\Gamma(1+\veps)\Big[{1\over\veps}\ln\left({t\over u}\right)\nn\\
&+&{1\over2}\ln^2\left({\mu^2\over -u}\right)-{1\over2}\ln^2\left({\mu^2\over -t}\right)\Big]\Bigg\},
\label{eq:qqb-qcdbox}
\end{eqnarray}
\begin{eqnarray}
\sumqn\left|\M^{\Box{(\qqb\ra\bbb)}}_{\rm EW}\right|^2 &=&4\*\pi\alpha\*\alpha_s^2\*{N^2-1\over N^2}\nn\\
&\times&\Bigg\{\gvq\gvb\Bigg[2\*\ln\left({t\over u}\right)\nn\\
&-&2\*z\*\bigg[\ln\left({tu\over\mz^4}\right)-2\*{s-\mz^2\over s}\*\ln\left(\left|1-{s\over\mz^2}\right|\right)-4\*\Li2{1-{s\over\mz^2}}\bigg]\nn\\
&-&4\*\bigg[\Big({s\*(1+z^2)\over s-\mz^2}-{\mz^2\over s}\Big)\*\ln\left(1-{s\over\mz^2}\right)\ln\left({t\over u}\right)\nn\\
&-&z\*\ln\left(1-{s\over\mz^2}\right)\ln\left({s^2\over tu}\right)\bigg] \nn\\
&-&2\*\bigg[\Big({s\*(1+z^2)\over s-\mz^2}-{2\*\mz^2\over s}\Big)\*\left(\Li2{1+{t\over\mz^2}}-\Li2{1+{u\over\mz^2}}\right)\nn\\
&+&2\*z\*\left(\Li2{1+{u\over\mz^2}}+\Li2{1+{t\over\mz^2}}\right)\bigg]\Bigg]\nn\\
&+&\gab\*\gaq\Bigg[2\*\ln\left({tu\over\mz^4}\right)-2\*z\ln\left({t\over u}\right)
-4\*{s-\mz^2\over s}\*\ln\left(\left|1-{s\over\mz^2}\right|\right)\nn\\
&+&{4\mz^2\over s}\bigg[\Li2{1+{t\over\mz^2}}+\Li2{1+{u\over\mz^2}}-2\*\Li2{1-{s\over\mz^2}}\nn\\
&-&\ln\left(1-{s\over\mz^2}\right)\ln\left({s^2\over tu}\right)\bigg]\nn\\
&+&4\*z\*{s+\mz^2\over  s-\mz^2}\ln\left(1-{s\over\mz^2}\right)\ln\left({t\over u}\right)\nn\\
&+&{4\*\mz^2\over s-\mz^2}\*z\*\bigg[\Li2{1+{t\over\mz^2}}-\Li2{1+{u\over\mz^2}}\bigg]\Bigg]\Bigg\}\nn\\
&+&\pi\*\alpha\*\alpha_s^2\*{N^2-1\over  N^2}\*\BB^{\qqb\ra\bbb}\*\Bigg\{(4\pi)^{\veps}\Gamma(1+\veps)\Big[{1\over\veps}\ln\left({t\over u}\right)\nn\\
&+&{1\over2}\ln^2\left({\mu^2\over -u}\right)-{1\over2}\ln^2\left({\mu^2\over -t}\right)\Big]\Bigg\},
\label{eq:qqb-ewbox}
\end{eqnarray}
with
\begin{eqnarray}
\BB^{\qqb\ra\bbb} &=&{8\*s\over s-\mz^2}\*\Big((d-3+z^2)\*\gvq\*\gvb-z\*(d-2)\*(d-3)\*\gaq\*\gab\Big).
\label{eq:qqb_born_QCDweak}
\end{eqnarray}
The factor $\BB^{\qqb\ra\bbb}$ results from the interference between the leading
order QCD amplitude and the weak amplitude. 
Terms proportional to the $Z$ vector couplings are odd functions in the
scattering angle $z$ as a consequence of Furry's theorem. 
The virtual corrections for the remaining quark-induced
processes can be deduced by the crossing relations \Eq{eq:quark-cross}.
For latter use we extract the IR divergent contribution to the differential cross
section
\begin{equation}
{\ind\sigma_{\qqb\ra\bbb}^{\Box\:\:\rm IR}\over \ind z} = 
{\alpha\*\alpha_s^2\over 32s}\*{N^2-1\over  N^2}\*\BB^{\qqb\ra\bbb}\*\Bigg\{(4\pi)^{\veps}\Gamma(1+\veps)\Big[{2\over\veps}\ln\left({t\over u}\right)
+\ln^2\left({\mu^2\over -u}\right)-\ln^2\left({\mu^2\over -t}\right)\Big]\Bigg\}.
\label{eq:virtual_IR_part}
\end{equation}
The contribution from real emission is given by
\begin{eqnarray}
\sumqn\left|\M^{\qqb\ra\bbb g}\right|^2 &=& 
\alpha_s^2\*\alpha\*(4\*\pi)^3\:\*{N^2-1\over N^2}\nn\\
&\times&\Big(\gvq\*\gvb\*(\tf^2+\ts^2+\uf^2+\us^2)-\gaq\*\gab\*(\tf^2+\ts^2-\uf^2-\us^2)\Big)\nn\\
&\times&{1\over s}\:\:\*{1\over s-\mz^2}\:\:\*{1\over s+\tf+\ts+\uf+\us}\:\:\*{1\over s+\tf+\ts+\uf+\us+\mz^2}\nn\\
&\times&{1\over s+\tf+\uf}\:\:\*{1\over s+\ts+\uf}\:\:\*{1\over s+\tf+\us}\:\:\*{1\over s+\ts+\us}\nn\\
&\times&\Big(2\*s^2 + (\tf+\ts+\uf+\us)\*(2\*s-\mz^2)\Big) \nn\\
&\times&\Big((\tf+\ts-\uf-\us)\*s^2+((\tf+\ts)^2-(\uf+\us)^2)\*s\nn\\
&+&(\tf+\ts+\uf+\us)\*(\tf\*\ts-\uf\*\us)\Big)\nn\\
\label{eq:real-qqb}
\end{eqnarray}
with
\begin{eqnarray}
\tf &=& -2\kq\kb, \quad \ts = -2\kqb\kbb,\nn\\
\uf &=& -2\kq\kbb, \quad \us = -2\kqb\kb.
\end{eqnarray}
The combination $s+\tf+\ts+\uf+\us+\mz^2$ appearing in \Eq{eq:real-qqb} can vanish, 
corresponding to on-shell production of the $Z$ boson. Therefore we introduced according to the 
principal value prescription a cut around the singular point
\begin{equation}
\delta > \sqrt{|s+\tf+\ts+\uf+\us+\mz^2|}.\nn
\end{equation}
and demonstrated numerically that the result remains unchanged for $\delta$ between $10^{-1}$ and $10^{-5}$ GeV. 
Moreover we checked that also a fictitious $Z$ boson width $0.01 \:\:{\rm GeV} \le \Gamma_Z \le 1\:\: {\rm GeV}$
produces the same numerical result. In analogy with the leading-order crossing relations we find for the
remaining real corrections:
\begin{equation}
\sumqn\left|\M^{q\bb\ra q\bb g}\right|^2 = \sumqn\left|\M^{\qb b\ra
  \qb b g}\right|^2 = \sumqn\left|\M^{\qqb\ra \bbb
  g}\right|^2\Bigg|_{s\lra t_1,\:\: t_2\ra -s-t_1-t_2-u_1-u_2}, 
\end{equation}
\begin{eqnarray}
\sumqn\left|\M^{qb\ra qb g}\right|^2 &=& \sumqn\left|\M^{\qb \bb\ra \qb \bb g}\right|^2 \nn\\
&=& \sumqn\left|\M^{q\bb\ra q\bb g}\right|^2\Big|_{s\lra u_2,\:\: u_1\ra -s-t_1-t_2-u_2-u_1}\nn\\
&=& \sumqn\left|\M^{\qqb\ra \bbb
  g}\right|^2\Bigg|_{s\ra t_1,\:\: t_1 \ra u_1,\:\: t_2 \ra u_2,\:\: u_1 \ra s,\:\: u_2\ra -s-t_1-t_2-u_1-u_2}.\nn\\
\end{eqnarray}
For these contributions the $Z$ boson remains off-shell.
To get an infrared finite result the corresponding subtraction terms from
the dipole formalism were implemented. We checked explicitly the
numerical stability of real corrections and dipoles and found the pointwise cancellation between the two contributions
in the singular phase space regions.
However the dipoles approximate the real corrections very well also in non-singular regions. This results in large cancellations in
the integrated result. In contrast to the singular configurations
this cancellation is not pointwise but takes place between different
phase space regions. As a consequence of this behaviour one needs high
statistics for the numerical integration. To check the numerical results we
therefore implemented also a variant of the phase-space-slicing method \cite{Beenakker:1988bq} and find
agreement comparing the results of both methods. The relevant formulae for the slicing method
and the plot for the comparison are shown in the Appendix \ref{sec:slicing}.\\

Let us now come to the contributions from the dipole formalism 
${\bf F}\otimes \ind\sigma^{\BB}\nn$ with ${\bf F} ={\bf I,K,P}$. 
The symbol `$\otimes$' denotes spin and colour
correlation between the operator ${\bf F}$ and the leading order
amplitudes, which are treated in general as vectors in colour and spin space. For the
processes under consideration, the gluon is always emitted from a fermion line and only trivial spin
correlation appears. We start with the contribution for the quark--antiquark annihilation
channel 
\begin{eqnarray}
\left({\bf I}\otimes \ind\sigma^{\BB}\right)^{\qqb\ra\bbb} &=& -{\alpha\alpha_s^2\over32s}\*{N^2-1\over N^2}{\BB^{\qqb\ra\bbb}}\Bigg\{{(4\pi)^{\veps}\over\Gamma(1-\veps)}
\*\Big[{2\over\veps}\*\ln\left({t\over u}\right)\nn\\
&+&\ln^2\left({\mu^2\over-u}\right)-\ln^2\left({\mu^2\over-t}\right) +
3\ln\left({t\over u}\right)\Big]\Bigg\},\nn\\
\label{eq:Iop_qqb}
\end{eqnarray}
where ${\BB^{\qqb\ra\bbb}}$ is defined in \Eq{eq:qqb_born_QCDweak}.
This contribution to the differential cross section cancells
the IR-poles and the $\mu$-dependence from the virtual corrections in
\Eq{eq:virtual_IR_part}.
For the finite ${\bf K}$ and ${\bf P}$ parts we find
\begin{eqnarray}
\left({\bf K}\otimes \ind\sigma^{\BB}\right)^{\qqb\ra\bbb}(x) &=& 0,\\
\label{eq:P_op_qqb}
\left({\bf P}\otimes \ind\sigma^{\BB}\right)^{\qqb\ra\bbb}(x) &=&
{\alpha\alpha_s^2\over 32sx}\*{N^2-1\over N^2}{\BB^{\qqb\ra\bbb}}\*\ln\left({t\over  u}\right)\left({1+x^2\over1-x}\right)_+.
\end{eqnarray} Of course in \Eq{eq:P_op_qqb}
${\BB^{\qqb\ra\bbb}}$ is also a function of $x$ and for the evaluation
of the plus distribution the well known relation 
\begin{equation}
\int_0^1 \ind x \Big(f(x)\Big)_+ g(x) = \int^1_0 \ind x \:f(x)\: [g(x)-g(1)]
\end{equation}
is used. 
For $\bb q\ra \bb q$ and $b \qb\ra  b \qb$ we find
\begin{eqnarray}
\left({\bf I}\otimes \ind\sigma^{\BB}\right)^{\bb q\ra\bb q}&=&\left({\bf I}\otimes \ind\sigma^{\BB}\right)^{b \qb\ra b \qb} \nn\\
&=&-{\alpha\alpha_s^2\over32s}\*{N^2-1\over N^2}{\BB^{\bb q\ra\bb q}}\Bigg\{{(4\pi)^{\veps}\over\Gamma(1-\veps)}
\*\Big[{2\over\veps}\*\ln\left({s\over -u}\right)\nn\\
&+&\ln^2\left({\mu^2\over-u}\right)-\ln^2\left({\mu^2\over s}\right)+3\ln{\left({s\over -u}\right)}\Big]\Bigg\},\nn\\
\label{eq:Iop_qbb}
\end{eqnarray}
\begin{eqnarray}
\left({\bf K}\otimes \ind\sigma^{\BB}\right)^{\bb q\ra\bb q}(x) &=& \left({\bf K}\otimes \ind\sigma^{\BB}\right)^{b \qb\ra b \qb}(x) \nn\\
&=&{\alpha\alpha_s^2\over 32sx}\*{N^2-1\over N^2}{\BB^{\bb q\ra\bb q}}
\*\Bigg[\delta(1-x)\*\left({3\over2}-{\pi^2\over3}\right)-(1+x)\*\ln(1-x)\nn\\
&+&{3\over2}\left({1\over1-x}\right)_++\left({2\ln(1-x)\over1-x}\right)_+\Bigg],
\end{eqnarray}
\begin{eqnarray}
\left({\bf P}\otimes \ind\sigma^{\BB}\right)^{\bb q\ra\bb q}(x) &=& \left({\bf P}\otimes \ind\sigma^{\BB}\right)^{b \qb\ra b \qb}(x) \nn\\
&=&-{\alpha\alpha_s^2\over 32sx}\*{N^2-1\over N^2}{\BB^{\bb q\ra\bb q}}\*\ln\left({-u\over  s}\right)\left({1+x^2\over1-x}\right)_+,
\end{eqnarray}
where $\BB^{\bb q\ra\bb q}$ can be deduced from $\BB^{\qqb\ra\bbb }$ and \Eq{eq:quark-cross}.
For $\bb \qb\ra \bb \qb$ (and $bq\ra  bq$) we find
\begin{eqnarray}
\left({\bf I}\otimes \ind\sigma^{\BB}\right)^{\bb \qb\ra\bb \qb}&=&\left({\bf I}\otimes \ind\sigma^{\BB}\right)^{b q\ra b q} \nn\\
&=&{\alpha\alpha_s^2\over32s}\*{N^2-1\over N^2}{\BB^{\bb \qb\ra\bb \qb}}\Bigg\{{(4\pi)^{\veps}\over\Gamma(1-\veps)}
\*\Big[{2\over\veps}\*\ln\left({s\over -u}\right)\nn\\
&+&\ln^2\left({\mu^2\over-u}\right)-\ln^2\left({\mu^2\over s}\right)+3\ln{\left({s\over -u}\right)}\Big]\Bigg\},\nn\\
\label{eq:Iop_qb}
\end{eqnarray}
\begin{eqnarray}
\left({\bf K}\otimes \ind\sigma^{\BB}\right)^{\bb \qb\ra\bb \qb}(x) &=& \left({\bf K}\otimes \ind\sigma^{\BB}\right)^{b q\ra b q}(x) \nn\\
&=&-{\alpha\alpha_s^2\over 32sx}\*{N^2-1\over N^2}{\BB^{\bb \qb\ra\bb \qb}}
\*\Bigg[\delta(1-x)\*\left({3\over2}-{\pi^2\over3}\right)-(1+x)\*\ln(1-x)\nn\\
&+&{3\over2}\left({1\over1-x}\right)_++\left({2\ln(1-x)\over1-x}\right)_+\Bigg],
\end{eqnarray}
\begin{eqnarray}
\left({\bf P}\otimes \ind\sigma^{\BB}\right)^{\bb \qb\ra\bb \qb}(x) &=& \left({\bf P}\otimes \ind\sigma^{\BB}\right)^{b q\ra b q}(x) \nn\\
&=&{\alpha\alpha_s^2\over 32sx}\*{N^2-1\over N^2}{\BB^{\bb \qb\ra\bb \qb}}\*\ln\left({-u\over  s}\right)\left({1+x^2\over1-x}\right)_+,
\end{eqnarray}
and $\BB^{\bb \qb\ra\bb \qb}$ is obtained by the crossing relations
\Eq{eq:quark-cross} as above. 
\subsection{Weak corrections to gluon-induced processes}
\label{sec:g-NLO} 
\begin{figure}[!tbp]
  \begin{center}
    \leavevmode
    \includegraphics[width=4.7cm]{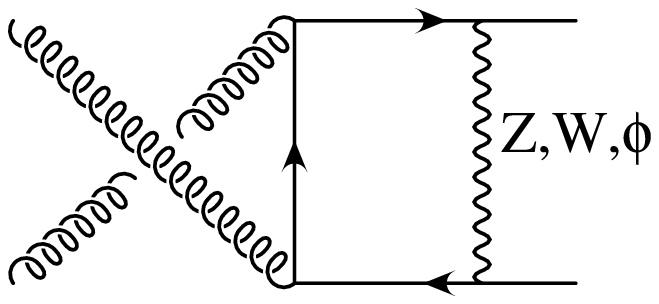}
    \hspace*{0.1cm}
    \includegraphics[width=4.cm]{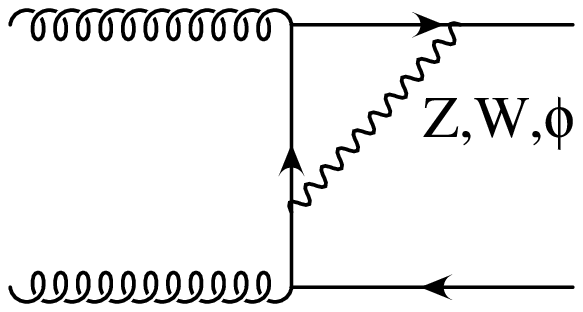}
    \hspace*{0.4cm}
    \includegraphics[width=4.cm]{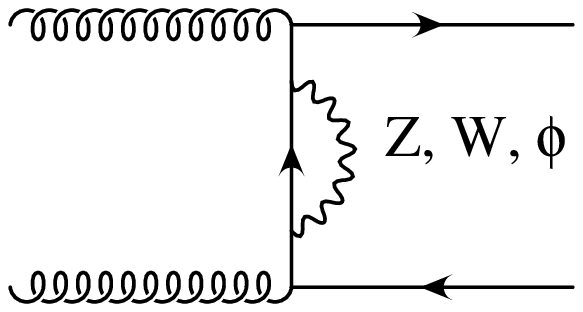}
    \caption{Sample diagrams for the gluon fusion process at $\order{\alpha_s^2\alpha}$.}
    \label{fig:ggNLO}
  \end{center}
\rput(0.1,3.75){{\large (a)}}
\rput(5.15,3.75){{\large (b)}}
\rput(9.8,3.75){{\large (c)}}
\end{figure}

Sample diagrams for the virtual corrections to the gluon-induced processes are shown in Fig~\ref{fig:ggNLO}.
No infrared divergencies are present and no corrections from real radiation contribute. 
The weak corrections are again classified into those from self-energy-, vertex- and
box-diagrams. The latter are UV finite, vertex and self-energy contributions must be renormalized. 
The differential cross section for the gluon fusion channel at next-to-leading order is 
decomposed as follows:
\begin{eqnarray}
  \label{eq:ggDecomposition}
  {\ind\sigma^{\order{\alpha_s^2\alpha}gg\ra\bbb}\over \ind z} &=& {1\over32\pi s}
  \sum_{i=Z,W,\phi}  \Bigg[\sumqn|\M^{\Box_i(gg\ra\bbb)}|^2+\sumqn|\M^{V_i(gg\ra\bbb)}|^2\nn\\
&+&\sumqn|\M^{\Sigma_i(gg\ra\bbb)}|^2 \Bigg].
\end{eqnarray}
We start with the analytic results for the self-energy diagrams (Fig~\ref{fig:ggNLO}(c)) in the
$gg\ra\bbb$ channel, which can be written in a compact form
\begin{eqnarray}
\label{eq:ggZself}
\sumqn|\M^{\Sigma_Z(gg\ra\bbb)}|^2 &=& \pi\*\alpha_s^2\alpha\*\Bigg\{\facselfz\*
\Big[ 3\*s\*(1+z)+4\*\mz^2\nn\\
&-&{2\over s\*(1+z)}\*\Big(s\*(1+z)+2\*\mz^2\Big)^2
\*\ln\left(1+{s\*(1+z)\over2\*\mz^2}\right)\Big]+(z\ra-z)\Bigg\},\nn\\
\end{eqnarray}
\begin{eqnarray}
\label{eq:ggWself}
\sumqn|\M^{\Sigma_W(gg\ra\bbb)}|^2 &=& 4\*\pi\*\alpha_s^2\alpha\*\Bigg\{\*\facselfw\*\nn\\
&\times&\Bigg[{1\over2\*(\mt^2-\mw^2)}\*\Big(s\*(1+z)\*(\mt^2-3\*\mw^2)-4\*(\mt^2-\mw^2)^2\Big)\nn\\
&+&{\mw^2\over(\mt^2-\mw^2)^2}\*\Big(s\*(1+z)\*(2\*\mt^2-\mw^2)\nn\\
&-&2\*(\mt^2-\mw^2)^2\Big)\*\ln\left({\mt^2\over\mw^2}\right)\nn\\
&+&\left(s\*(1+z)+2\mw^2-2\mt^2\right)\*f(y_1,y_2)\Bigg]+(z\ra-z)\Bigg\},
\end{eqnarray}
where
\begin{eqnarray}
f(y_1,y_2) = -y_1\*\ln\L{y_1\over y_1-1}\R-y_2\*\ln\L{y_2\over y_2-1}\R
\end{eqnarray}
\begin{eqnarray}
y_{1/2} &=& {t+\mw^2-\mt^2\over2t}\*\times\left(1\pm\sqrt{1-{4\mw^2t\over
  (t+\mw^2-\mt^2)^2}}\right).\nn\\
\end{eqnarray}
The contributions involving $\phi$ and $W$ are closely related
\begin{equation}
\sumqn|\M^{\Sigma_\phi(gg\ra\bbb)}|^2 = {\mt^2\over2\*\mw^2}\*\sumqn|\M^{\Sigma_W(gg\ra\bbb)}|^2.
\end{equation}
The vertex corrections consist of $s$, $t$ and $u$ channel contributions. Since the corrections to the $s$ 
channel are proportional to the quark mass (see Ref.~\cite{Kuhn:2006vh} Eqs.(II.18-22)) they vanish in the
massless limit. The remaining vertex corrections (e.g. Fig.~\ref{fig:ggNLO}(b)) can be expressed in terms of the self
energies
\begin{equation}
\sumqn|\M^{V_i(gg\ra\bbb)}|^2=-2\sumqn|\M^{\Sigma_i(gg\ra\bbb)}|^2.
\end{equation}
The box contributions (Fig~\ref{fig:ggNLO}(a)) are more involved and are listed in Appendix \ref{sec:gg-boxes}.
Corrections to $gb\ra gb$ and $g\bb\ra g\bb$ are obtained via the relations \Eq{eq:gluon-cross}. 
\section{Results}
\label{sec:results}
\begin{figure}[!htbp]
  \begin{center}
    \leavevmode
    \includegraphics[width=12cm]{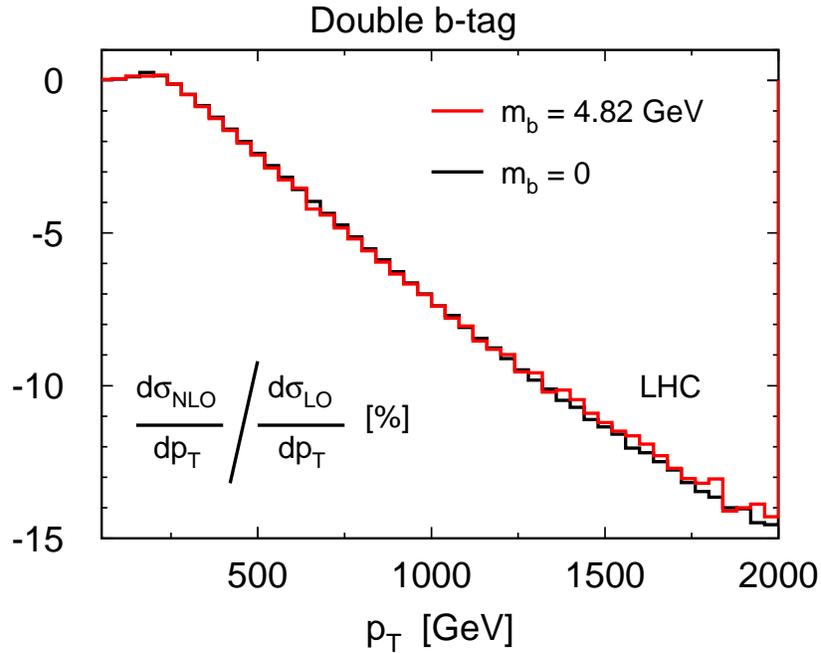}
     \caption{Relative weak corrections for double $b$-tag events at the LHC, comparing the results for a massless
       bottom-quark with the massive case.}
     \label{fig:pt-nlo-bjet-mass-compare}
  \end{center}
\end{figure}

\begin{figure}[!htbp]
  \begin{center}
    \leavevmode
    \includegraphics[width=12cm]{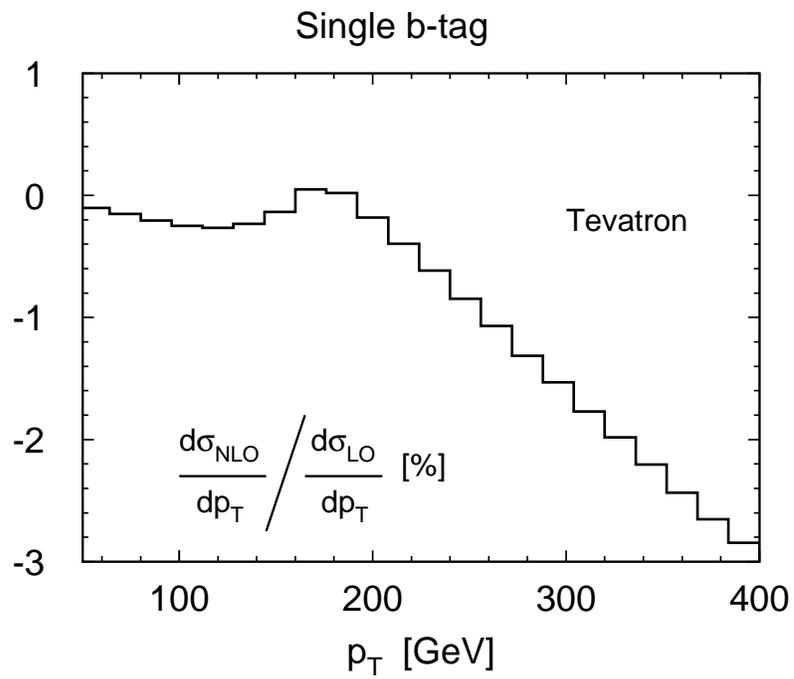}
    \includegraphics[width=12cm]{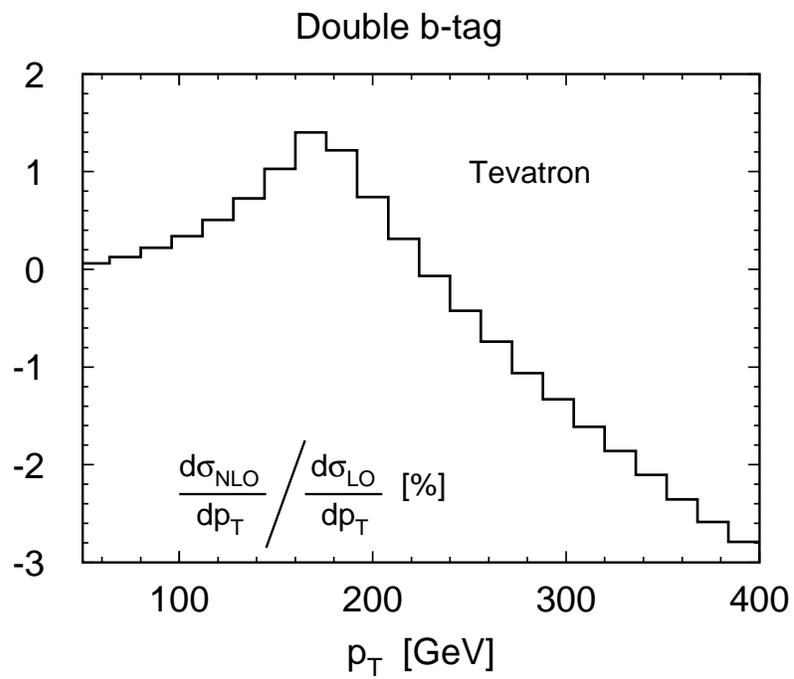}
     \caption{Relative weak corrections for single $b$-tag (upper figure) and 
       double $b$-tag events (lower figure) at the Tevatron.}
     \label{fig:pt-nlo-bjet-relative-TEV}
  \end{center}
\end{figure}

Apart of the obvious checks, like cancelation of UV and IR singularities,
a numerical test was performed for the
IR-divergent contributions, implementing the
phase-space-slicing method. As illustrated in Appendix B we found complete agreement of the two methods. 
For the quark-antiquark annihilation and
the gluon fusion channel we found complete analytical agreement with previous results for top-quark
production \cite{Kuhn:2006vh}, 
with the following replacements 
\begin{equation}
\mt\lra\mb, \quad \gvt\ra \gvb, \quad \gat\ra \gab \quad  {\rm and}
\quad {\mt\ra 0}.\nn
\end{equation}
These results also allow for a numerical comparison between the massive and the massless approach. 
In Fig.~\ref{fig:pt-nlo-bjet-mass-compare} the results for 
a massive bottom-quark ($\mb = 4.82$ GeV) are compared with those from the massless
approximation. Agreement between the massive and the massless results is observed in the full kinematic range, 
which justifies the massless approximation for the bottom-quarks used in the paper.
The difference of the massless and the massive result concerning the relative corrections is always below $0.5\%$.  
Furthermore all LO and NLO partonic corrections were calculated 
analytically and thus all crossing relations listed in the previous
Sections are used as additional cross check. For the numerical evaluation we use the same input parameters as in Section~\ref{sec:leadingorder}.
As mentioned in Section~\ref{sec:NLO} the results are 
leading order in QCD and therefore evaluated with the leading order PDF's CTEQ6L.\\
We start with the results for the differential $\pt$-distribution at the
Tevatron. In Fig.~\ref{fig:pt-nlo-bjet-relative-TEV} the relative
corrections normalised to the full leading order cross section are shown.
\begin{figure}[!htbp]
  \begin{center}
    \leavevmode
    \includegraphics[width=12cm]{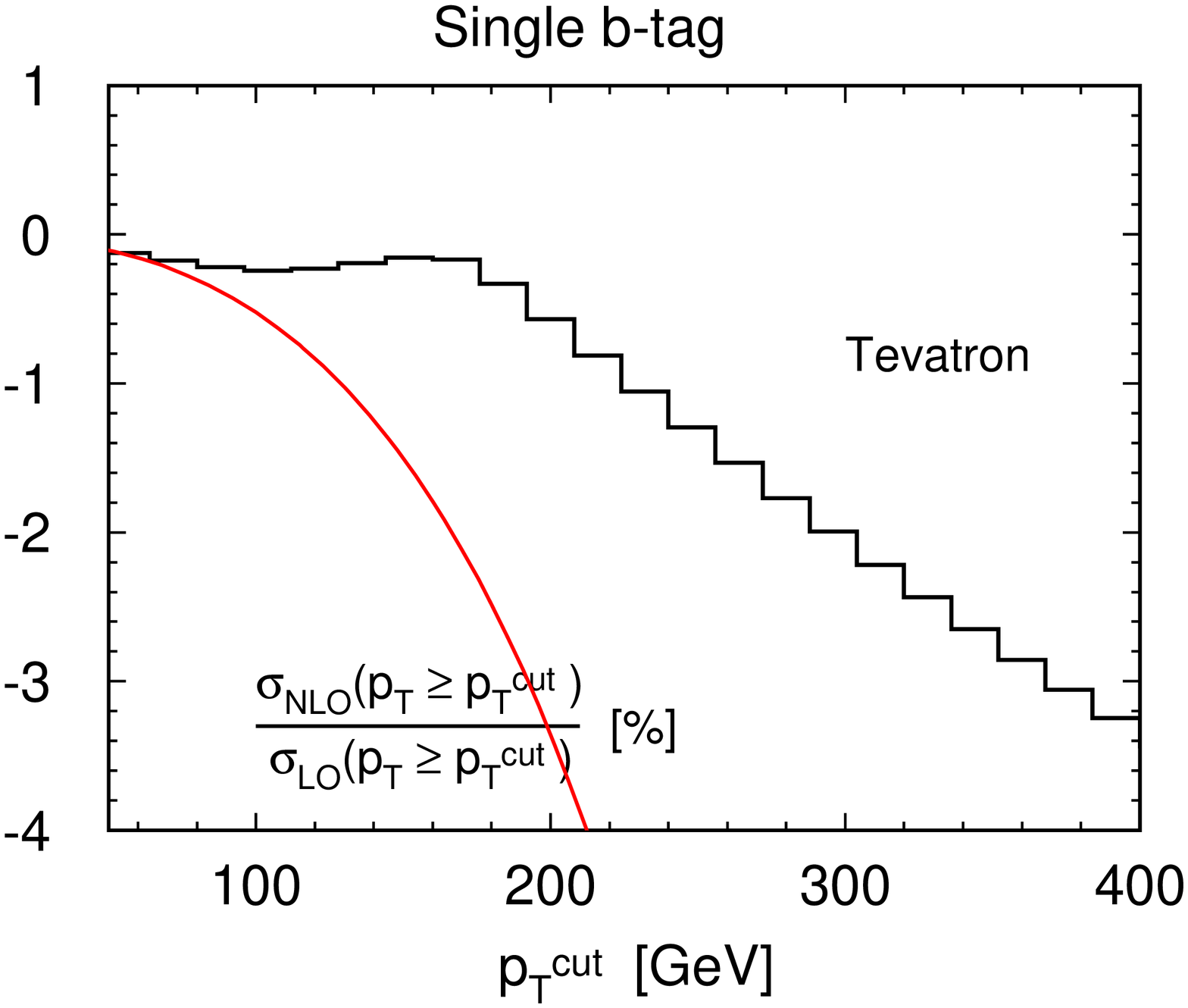}
    \includegraphics[width=12cm]{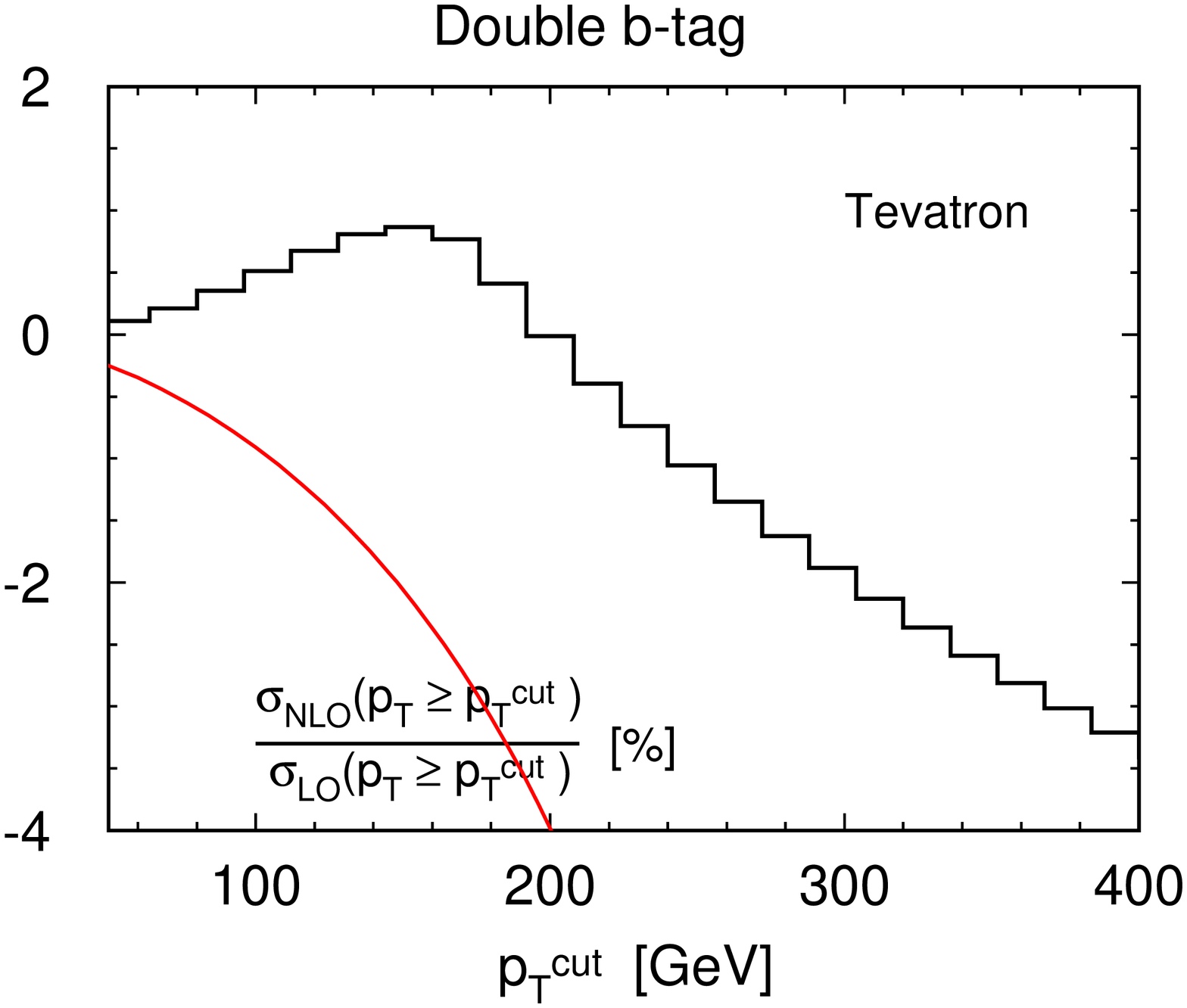}
     \caption{Relative corrections to the cross section
       for $\pt >\ptcut$ at the Tevatron for single $b$-tag (upper figure) and 
       double $b$-tag events (lower figure). The red lines give an estimate on
       the statistical uncertainty as described in the text.}
     \label{fig:pt-cut-nlo-bjet-relative-TEV}
  \end{center}
\end{figure}
For single $b$-tag events (upper figure) the relative corrections are negative and of the order of a few permille up to $\pt = 120$ GeV.
The following increase of the relative corrections with the resonant behaviour around $\pt = 180$ GeV is caused by the
vertex corrections to $\qqb\ra \bbb$ involving virtual top-quarks. In this range of $\pt$ the weak corrections become 
positive and increase the differential cross section by a few permille. 
Above the resonance peak the weak corrections become negative again and the typical Sudakov behaviour is observed. 
The magnitude of the relative corrections increases with energy for $\pt > 200$ GeV and reaches $-3\%$ for $\pt = 400$ GeV. \\
Due to the aforementioned virtual top-quarks, the relative corrections for double $b$-tag events are positive 
for $\pt$ values below 220 GeV (lower figure). 
In this case $\qqb\ra \bbb$ and $gg\ra \bbb$ are the only contributing partonic channels, hence the resonance peak is more distinct than in 
the single $b$-tag case with the maximum ($+1.5\%$) of the relative corrections around $\pt = 180$ GeV. The influence of the Sudakov logarithms starts for 
$\pt > 220$ GeV, where the weak corrections change sign and become negative. For $\pt = 400$ GeV the relative corrections amount up to $-3\%$. \\
The relative corrections to the integrated cross section ($\pt> \ptcut$) at the Tevatron are shown in Fig.~\ref{fig:pt-cut-nlo-bjet-relative-TEV}. 
For the single $b$-tag (upper figure) and the double $b$-tag (lower figure) 
the peak around the $t\bar{t}$-threshold is smoothed by contributions from higher $\pt$-values, leaving the integrated correction negative in the single $b$-tag case. Beyond 
$\pt = 200$ GeV the relative corrections to the integrated distributions are between $0$ and $-3\%$. In addition we give in Fig.~\ref{fig:pt-cut-nlo-bjet-relative-TEV}
a rough estimate of the expected statistical error at the Tevatron. The estimated number of events for $\pt> \ptcut$ is based on an integrated luminosity of 8
${\rm fb}^{-1}$. A comparison of the statistical uncertainty with the relative weak corrections shows that the weak effects will not be visible at the
Tevatron. 
 \\

\begin{figure}[!htbp]
  \begin{center}
    \leavevmode
    \includegraphics[width=12cm]{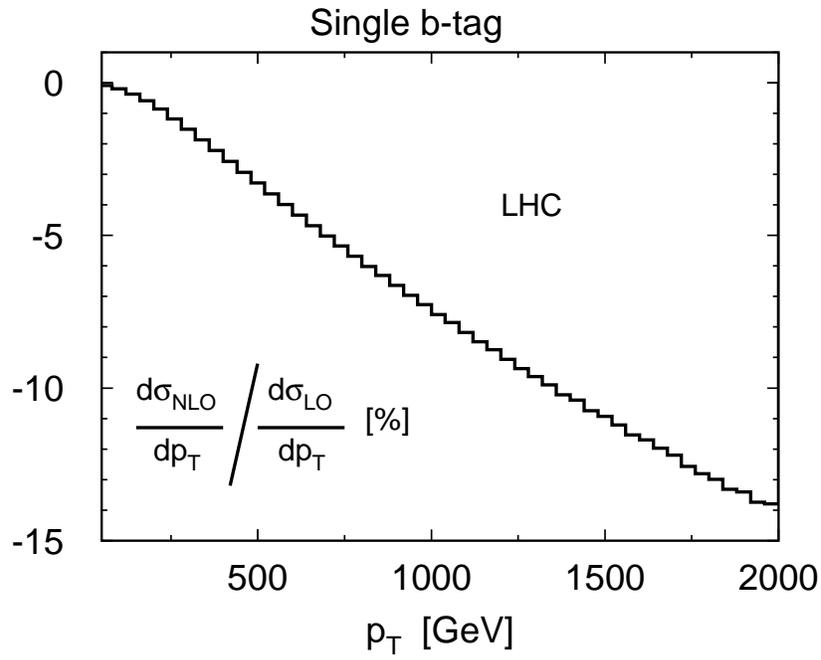}
    \includegraphics[width=12cm]{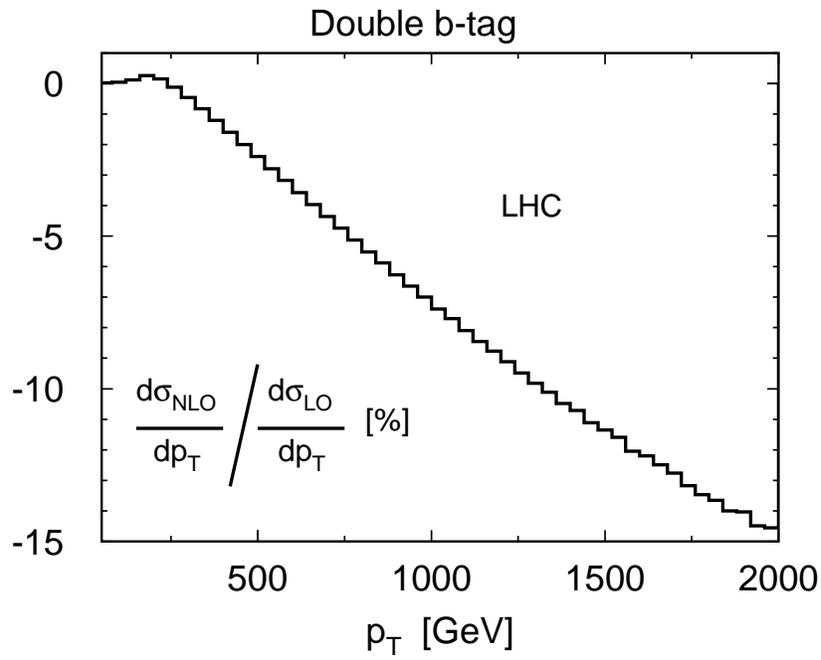}
     \caption{Relative weak corrections for single $b$-tag (upper figure) and 
       double $b$-tag (lower figure) events at the LHC.}
     \label{fig:pt-nlo-bjet-relative-LHC}
  \end{center}
\end{figure}

\begin{figure}[!htbp]
  \begin{center}
    \leavevmode
    \includegraphics[width=12cm]{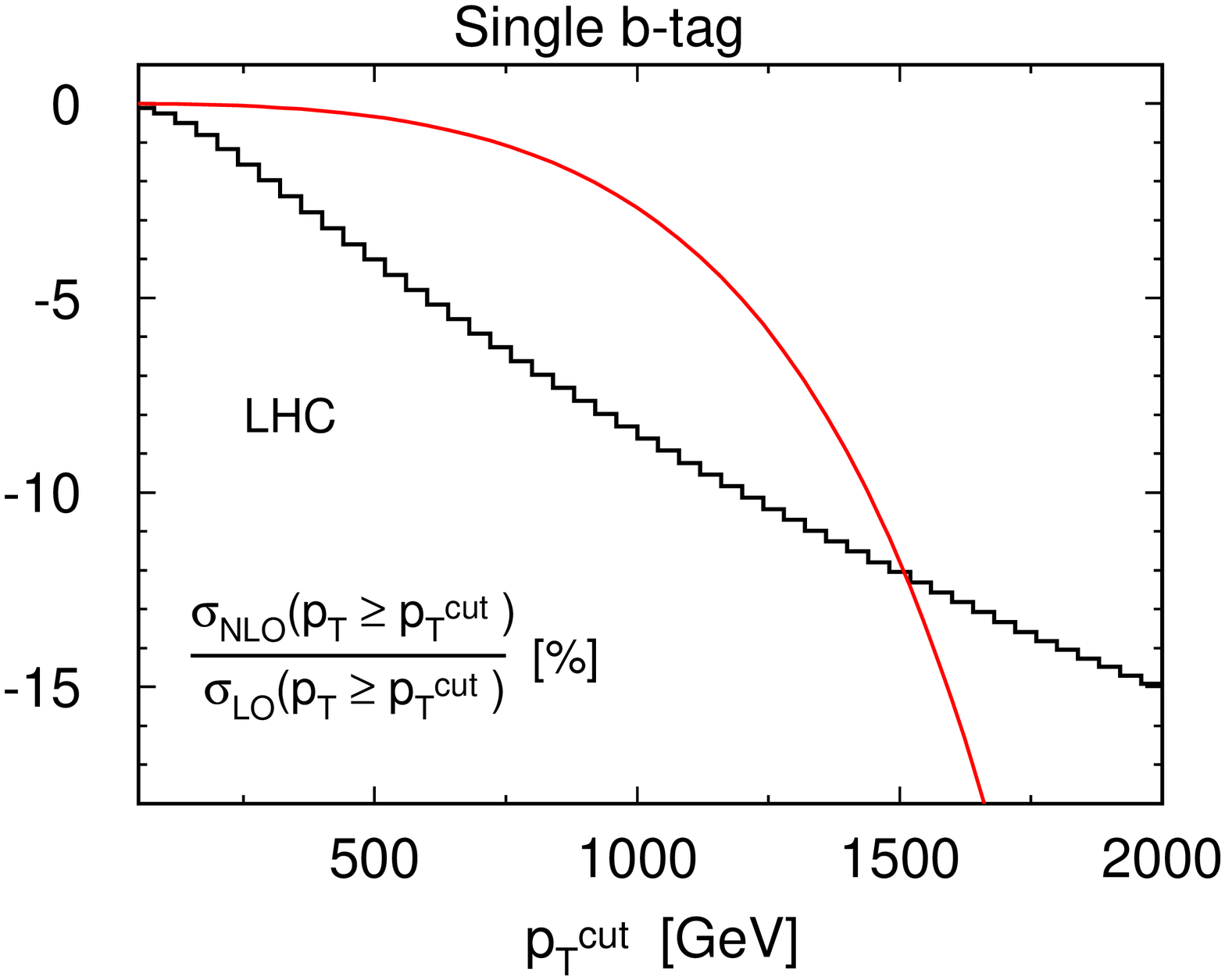}
    \includegraphics[width=12cm]{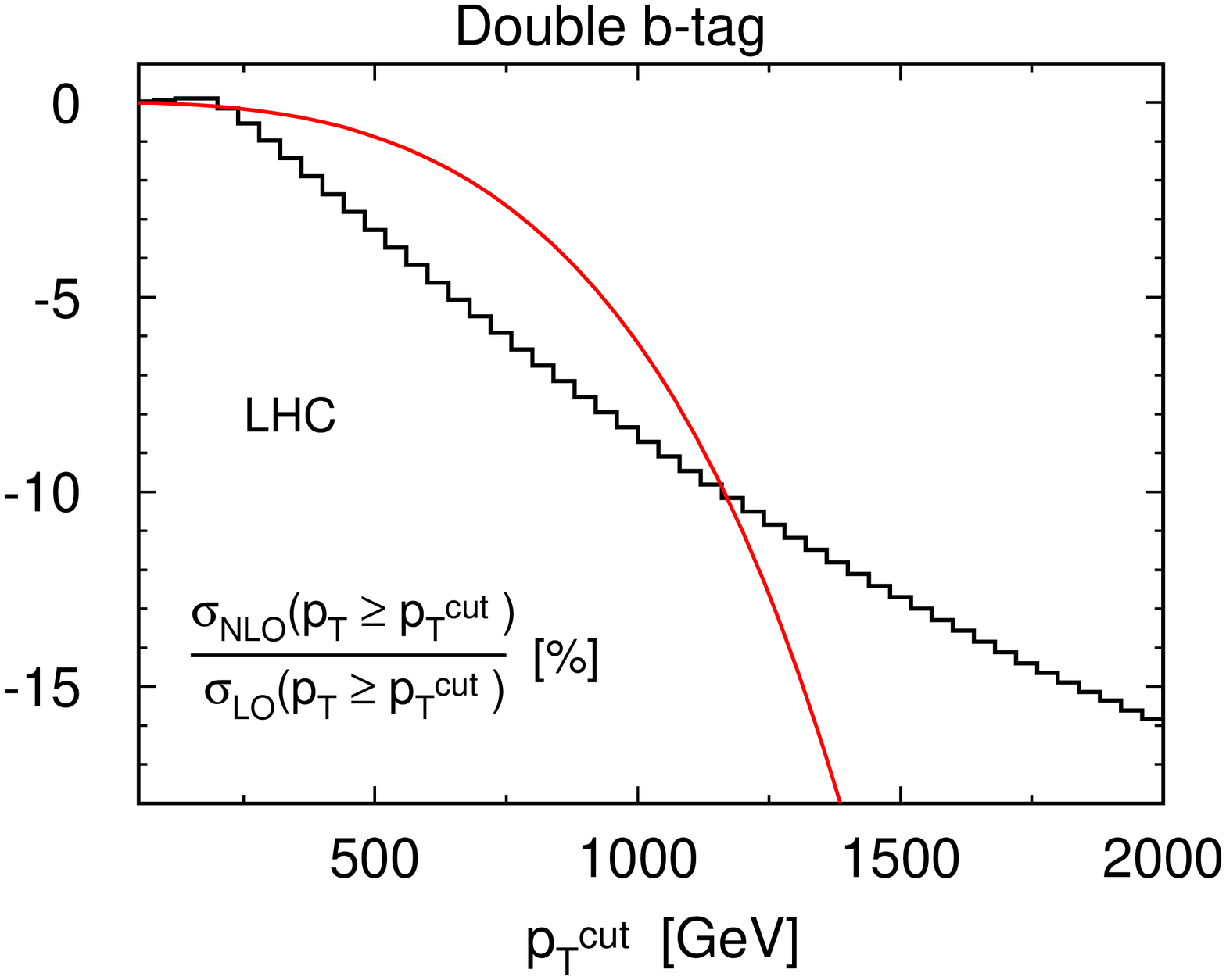}
     \caption{Relative corrections to the cross section for $\pt > \ptcut$ at the LHC for single $b$-tag (upper figure) and 
double $b$-tag events (lower figure). The red lines give an estimate on the statistical uncertainty as described in the text.}
     \label{fig:pt-cut-nlo-bjet-relative-LHC}
  \end{center}
\end{figure}

\begin{figure}[!htbp]
  \begin{center}
    \leavevmode
    \includegraphics[width=0.49\textwidth]{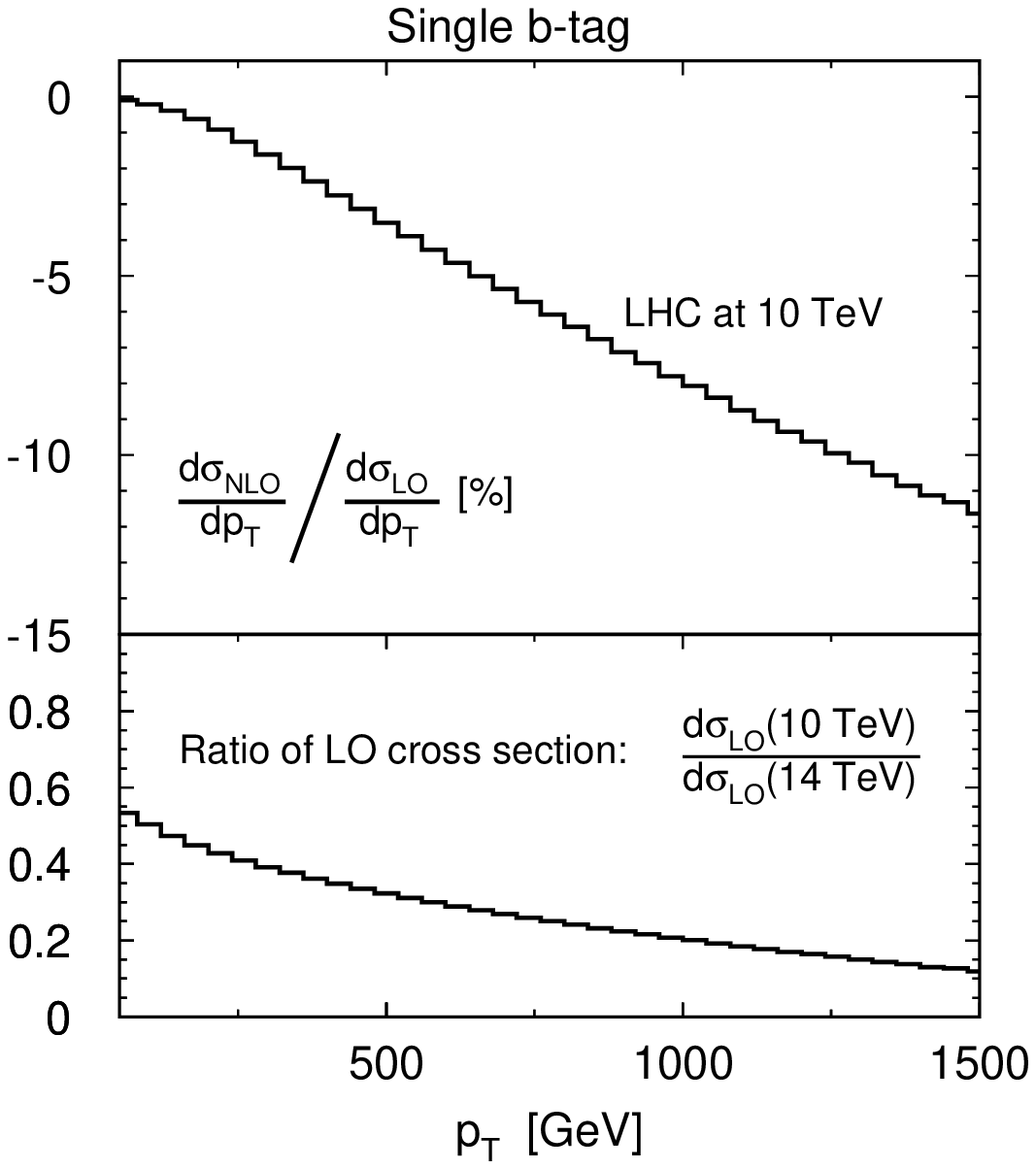}
    \includegraphics[width=0.49\textwidth]{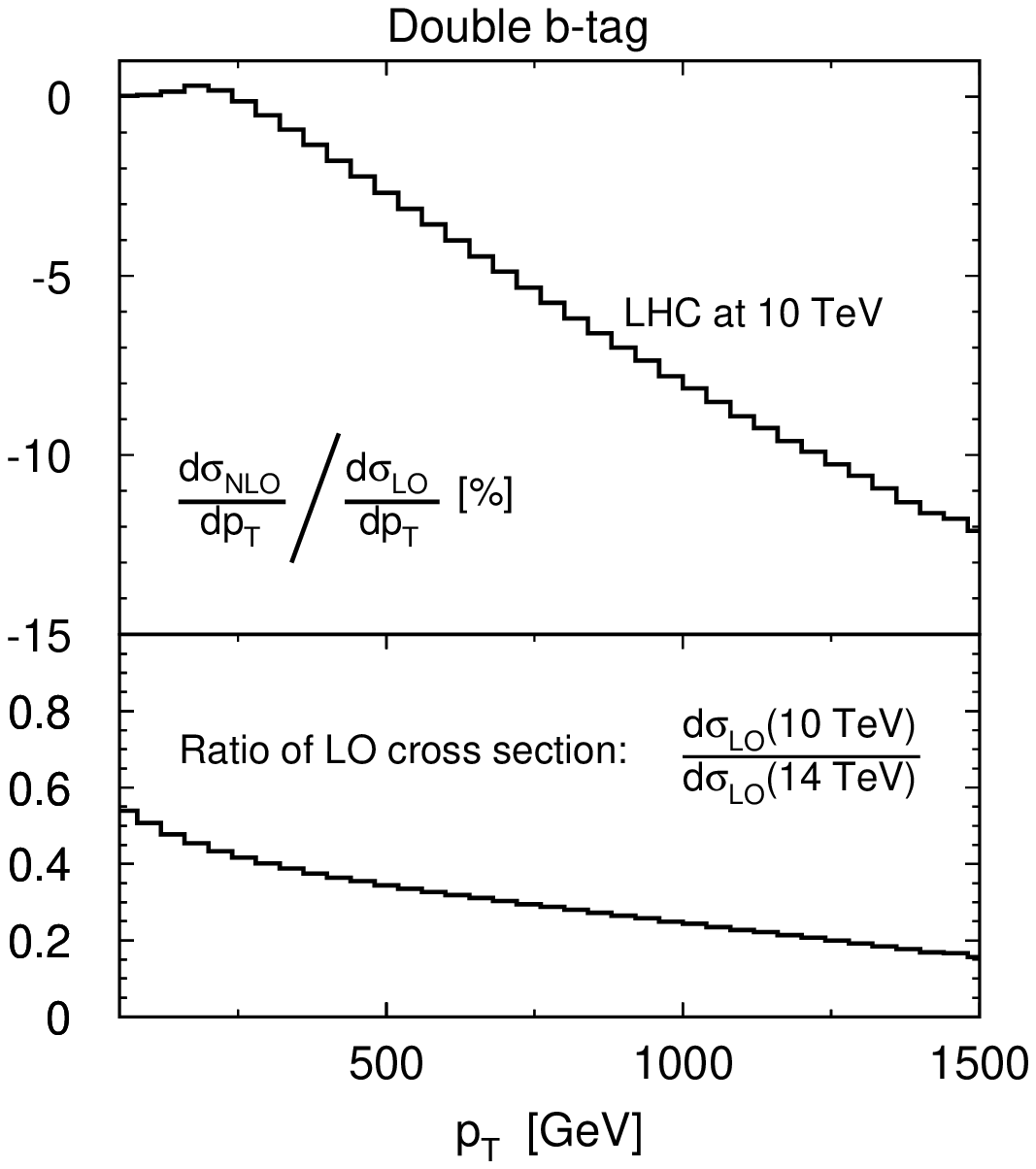}
     \caption{Relative corrections for single $b$-tag (a) and double $b$-tag events (b) at the LHC operating with a center-of-mass energy $\sqrt{s} = 10$ TeV (upper figures) 
       and the ratio of the differential leading order cross sections at the LHC operating at 10 TeV and 14 TeV 
       $\left(\ind\sigma_{\rm LO}(10\:{\rm TeV})/ \ind\pt\right)/\left(\ind\sigma_{\rm LO}(14\:{\rm TeV})/ \ind\pt\right)$ (lower figures).}
     \label{fig:pt-LHC10}
  \end{center}
  \rput(6.2,10.45){\large (a)}
  \rput(13.4,10.45){\large (b)}
\end{figure}

At the LHC however the corrections are significantly larger. For the subsequent discsusion $\sqrt{s} = 14$ TeV will be adopted.
The relative corrections to the leading order $\pt$-distributions are shown in Fig.~\ref{fig:pt-nlo-bjet-relative-LHC}.
For single $b$-tag events (upper figure) the relative corrections are always negative and of the order of a few permille up to $-1\%$ for 50 GeV $< \pt <$ 250 GeV. 
In contrast to the Tevatron results, the resonant behaviour arising from virtual top-quarks in quark--anti-quark annihilation is not visible, 
a consequence of the dominance of the gluon-induced processes in this $\pt$-regime (Fig.~\ref{fig:pt-bjet-lhc}(a)).
For 250 GeV $< \pt< $ 1 TeV the weak NLO contributions vary between $-1\%$ and $-8\%$, compared to the leading order distribution, a consequence of the Sudakov logarithms.
In the high energy regime ($\pt> 1$ TeV) the relative corrections amount to $-10\%$ and will even reach $-14\%$ for $\pt = 2$ TeV. \\
The lower figure in Fig.~\ref{fig:pt-nlo-bjet-relative-LHC} shows the relative corrections for double $b$-tag events at the LHC. 
Despite the strong suppression of $\qqb\ra\bbb$ at leading order (see Fig.~\ref{fig:pt-bjet-lhc}(b)) a small remnant of the enhancement  
from virtual top-quarks is visible in the double $b$-tag case. 
With increasing $\pt$ the Sudakov logarithms dominate the shape of the weak NLO contributions 
and yield relative corrections between $-1\%$ and $-7\%$ (250 GeV $< \pt< $ 1 TeV). At the highest $\pt$-values considered relative corrections 
up to $-14\%$ are observed. 
\\
Fig.~\ref{fig:pt-cut-nlo-bjet-relative-LHC} (upper figure) shows the integrated cross section for single $b$-tag events at the LHC together with an estimate of the
statistical error based on an integrated luminosity of 200 ${\rm fb}^{-1}$. The same composition in shape
and magnitude is observed as for the differential distribution. The statistical error estimate matches the size of the weak corrections up to $\pt = 1.5$ TeV. 
For higher $\pt$-values the rate drops quickly and it will be difficult to observe the effect of the weak corrections. For double $b$-tag events (lower figure) 
we find again the smoothing of the $t\bar{t}$-threshold for the relative weak corrections, while the composition of the curve for $\pt> 250$ GeV is very similiar to
the already discussed differential distribution. Considering the statistical error, the slight increase from virtual top-quarks will not be observable at the LHC. 
For $\pt$-values between $250$-$1000$ GeV the weak corrections are larger than the statistical error, above $\pt =1$ TeV they are comparable 
or smaller.\\ 
In Fig.~\ref{fig:pt-LHC10} we show results for the LHC operating at $\sqrt{s} = 10$ TeV. 
The absolute cross sections are by more than a factor two lower, due to the lower parton luminosities.
However the impact of the electroweak corrections is nearly the same and qualitatively similiar results are obtained for the relative NLO corrections.
\section{Conclusion}
\label{sec:conclusion}
In this article we present the weak corrections for $b$-jet production
of order $\aas$ neglecting purely photonic corrections. We derive compact analytic results
for the NLO contributions in quark--anti-quark annihilation and the
gluon-fusion. For the remaining partonic processes we
list the crossing relations needed to derive the corresponding analytic 
results. We present the corrections to the
$\pt$-distribution and compare the results with the expected
statistical uncertainty. In particular we find corrections up to ten
percent for bottom-jets at high transverse momenta accessible at the
LHC. These effects are larger than the anticipated statistical uncertainty. \\

{\bf Acknowledgements:}
This work is supported by the European Community's Marie-Curie Research Training
Network under contract MRTN-CT-2006-035505 ``Tools and Precision Calculations for
Physics Discoveries at Colliders''. P.U. acknowledges the support of the
Initiative and Networking Fund of the Helmholtz Association, contract HA-101 
(``Physics at the Terascale'').
\newpage
\appendix
\section{Notation}
\label{sec:integrals}
We define as usual
\begin{eqnarray*}
  && {\rm A}_0(\m1^2) = {1\over i\pi^2} \int \ind^d\ell {(2\pi\mu)^{2\e}\over
  (\ell^2-\m1^2+i\e)}\\
\\
  &&{\rm B}_0(p_1^2,\m1^2,\m2^2)={1\over i\pi^2} \int \ind^d\ell 
  {(2\pi\mu)^{2\e}\over (\ell^2-\m1^2+i\e)
    ((\ell+p_1)^2-\m2^2+i\e)},\nn\\ 
\\
  &&{\rm C}_0(p_1^2,p_2^2,(p_1+p_2)^2,
  \m1^2,\m2^2,\m3^2)=\nn\\
  &&{1\over i\pi^2} \int \ind^d\ell 
  {(2\pi\mu)^{2\e}\over (\ell^2-\m1^2+i\e)
    ((\ell+p_1)^2-\m2^2+i\e)((\ell+p_1+p_2)^2-\m3^2+i\e)},\nn\\
\\
 &&{\rm D}_0(p_1^2,p_2^2,p_3^2,(p_1+p_2)^2,(p_1+p_3)^2,(p_2+p_3)^2,\m1^2,\m2^2,\m3^2,\m4^2 )=\nn\\
&&{1\over i\pi^2}\int{\ind^d\ell}{(2\pi\mu)^{2\e}\over
(\ell^2-\m1^2+i\e)((\ell+p_1)^2-\m2^2+i\e)((\ell+p_1+p_2)^2-\m3^2+i\e)}\\
&&\qquad\qquad\times{1\over((\ell+p_1+p_2+p_3)^2-\m4^2+i\e)}, \\
\\
 &&{\rm D}_0^6(p_1^2,p_2^2,p_3^2,(p_1+p_2)^2,(p_1+p_3)^2,(p_2+p_3)^2,\m1^2,\m2^2,\m3^2,\m4^2 )=\nn\\
&&{1\over i\pi^2}\int{\ind^6\ell}{1\over
(\ell^2-\m1^2+i\e)((\ell+p_1)^2-\m2^2+i\e)((\ell+p_1+p_2)^2-\m3^2+i\e)}\nn\\
&&\qquad\qquad\times{1\over((\ell+p_1+p_2+p_3)^2-\m4^2+i\e)}.
\end{eqnarray*}  
For the UV-divergent one-point integrals ${\rm A}_0$ and two-point integrals ${\rm B}_0$ we define the 
finite parts through
\begin{eqnarray}
{\rm A}_0(m^2) = m^2\Delta+{\overline{\rm A}}_0(m^2),\nn\\
{\rm B}_0(p^2,m_1^2,m_2^2) = \Delta+{\overline{\rm B}}_0(p^2,m_1^2,m_2^2),
\end{eqnarray}
with $\Delta = 1/\veps-\gamma+\ln{(4\pi)}$.
The 6-dimensional four point functions used in this article are 
\begin{eqnarray}
{\rm D}_0^6(0,0,0,s,u,t,0,0,0,0) &=& {\pi\over 2u}\ln^2\left({-t\over s}\right),\\
{\rm D}_0^6(0,0,0,s,u,t,0,0,0,\mz^2) &=& -{\pi\over
  su}\*\Bigg\{(\mz^2-s)\*\Big[\li\left(1-{s\over\mz^2}\right)\nn\\
&+&\ln{\left(\left|1-{s\over\mz^2}\right|\right)}\ln{\left(-{s\over t}\right)}\Big]\nn\\
&+& {s\*(t+\mz^2)\over t}\*\li{\left(1+{t\over\mz^2}\right)}-{\pi^2\over6}{\mz^2(s+t)\over t}\Bigg\}.\nn\\
\end{eqnarray}
The following abbreviations are used for 
the 4-dimensional four and three point function
\begin{eqnarray}
{\rm D}_0^W(z) &=& {\rm D}_0(0,0,0,t,u,s,\mt^2,\mt^2,\mw^2,\mt^2),
\end{eqnarray}
\begin{eqnarray}
{\rm C}_0(s,\mt^2,\mw^2,\mt^2) &=& {\rm C}_0(0,0,s,\mt^2,\mw^2,\mt^2), \nn\\
{\rm C}_0(s,\mt^2,\mt^2,\mt^2) &=& {\rm C}_0(0,0,s,\mt^2,\mt^2,\mt^2), \nn\\
{\rm C}_0 (t,\mt^2,\mt^2,\mw^2) &=& {\rm C}_0(0,0,t,\mt^2,\mt^2,\mw^2). 
\end{eqnarray}
\section{Phase space slicing}
\label{sec:slicing}
In this appendix we list the relevant formulae for the phase-space
slicing method applied to the quark-induced processes listed in Section~\ref{sec:q-NLO}. 
The integral over the three-particle phase space can be
written as
\begin{eqnarray}
&&\int {\ind^{d-1}k_1\over(2\pi)^3} {1\over 2E_1}\int
     {\ind^{d-1}k_2\over(2\pi)^3}{1\over 2E_2} \int {\ind^{d-1}k_g\over(2\pi)^3}
     {1\over 2E_g}\:\:\theta(E_c-E_g)  \left|\Mm\right|^2\nn\\
&+&\int {\ind^3k_1\over(2\pi)^3} {1\over 2E_1}\int
     {\ind^3k_2\over(2\pi)^3}{1\over 2E_2} \int {\ind^3k_g\over(2\pi)^3}
     {1\over 2E_g}\:\:\theta(E_g-E_c)\left|\Mm\right|^2 +\O(\veps).
\label{eq:3-particle-phase-in-d}
\end{eqnarray}
where we introduced a cut $E_c$ on the gluon energy $E_g$. The second
term in this equation is integrated in $d = 4$ dimensions, yielding
a logarithmic dependence of the result on the cut $E_c$. For the
first term the real matrix element squared is calculated in the
eikonal approximation and evaluated in $d$ dimensions, using 
the parametrisation given in \Ref{Beenakker:1988bq}. For the eikonal factor for the
$\qqb\ra\bbb$ process we find
\begin{eqnarray}
&\phantom{x}&\int {\ind^{d-1}k_g\over(2\pi)^3}{1\over 2E_g}\Bigg[{2\kq\cdot\kb\over2\kq\cdot\kg2\kb\cdot\kg}
+{2\kqb\cdot\kbb\over2\kqb\cdot\kg2\kbb\cdot\kg}
-{2\kq\cdot\kbb\over2\kq\cdot\kg2\kbb\cdot\kg}-{2\kqb\cdot\kb\over2\kqb\cdot\kg2\kb\cdot\kg}\Bigg]\nn\\
&=& {1\over8\pi^2}{1\over\veps^2}\pi^{\veps}{\Gamma(1-\veps)\over
  \Gamma(1-2\veps)}\left({\mu^2\over
  E_c^2}\right)^\veps\nn\\
&\times&\Bigg[\left({-t\over
      s}\right)^{-\veps}\left(1+\veps^2\Li2{{-u\over s}}\right)
-\left({-u\over s}\right)^{-\veps}\left(1+\veps^2\Li2{{-t\over s}}\right)
\Bigg]\nn\\
\label{eq:qqb-eikonal}
\end{eqnarray}
Adding the corresponding contribution from the virtual corrections \Eq{eq:virtual_IR_part} we find
\begin{eqnarray}
{\ind\sigma_{\qqb\ra\bbb}^{\rm Soft}\over \ind z} &=&
\Bigg[{\ind\sigma_{\qqb\ra\bbb}^{\Box\:\:\rm IR}\over \ind z}+{\ind\sigma_{\qqb\ra\bbb
      g}^{\rm Eikon}\over \ind z} \Bigg]_{\veps \ra 0}\nn\\
&=&{\aas\over 16s}\*{N^2-1\over
  N^2}\*\BB^{\qqb\ra\bbb}\*\Big[\ln{\left(x_{\rm min}^2\right)}\ln{\left({t\over u}\right)}-\Li2{-{t\over s}}+\Li2{-{u\over s}}\Big]\nn\\
\label{eq:qqb-soft}
\end{eqnarray}
with the dimensionless variable $x_{\rm min} \le {2E_g\over \sqrt{s}}$.
For the remaining partonic processes the "soft" contributions to the differential
cross section are given through
\begin{eqnarray}
{\ind\sigma_{q\bb\ra q\bb}^{\rm Soft}\over \ind z} &=&
{\aas\over 16s}\*{N^2-1\over N^2}\*\BB^{q\bb\ra q\bb}\*\Big[\ln{\left(x_{\rm min}^2\right)}\ln{\left(-{s\over u}\right)}-\Li2{-{t\over s}}+{\pi^2\over2}\Big],
\label{eq:qbb-soft}
\end{eqnarray}
\begin{eqnarray}
{\ind\sigma_{qb\ra qb}^{\rm Soft}\over \ind z} &=&
-{\aas\over 16s}\*{N^2-1\over N^2}\*\BB^{qb\ra qb}\*\Big[\ln{\left(x_{\rm min}^2\right)}\ln{\left(-{s\over u}\right)}-\Li2{-{t\over s}}+{\pi^2\over2}\Big].
\label{eq:qb-soft}
\end{eqnarray}
Finally we present the comparison between the phase space slicing and the dipol subtraction method.
In the figure below the difference for the differential cross sections at the LHC 
obtained with the two different methods in terms of standard deviations is shown. The agreement is always better than three sigma. 
\begin{figure}[!h]
  \begin{center}
    \includegraphics[width=12cm]{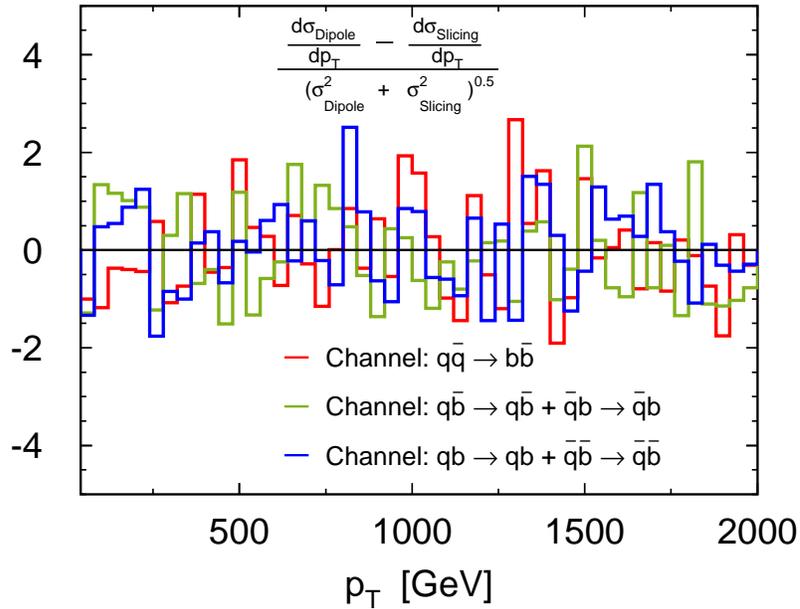}
     \caption{Difference between results based on dipole formalism and phase space slicing for the differential cross sections 
       for the different quark-induced channels in terms of standard deviations.
     For the comparison only the relevant IR contributions (boxes and real corrections) were taken into account.}
     \label{fig:cut-study}
  \end{center}
\end{figure}
\section{Box diagrams for gluon fusion}
\label{sec:gg-boxes}
\begin{eqnarray}
\sumqn\left|\Mm^{\Box_Z\:\:\aas}_{gg\ra\bbb}\right|^2 &=& 2\*\pi\*\aas\*{1\over s}\*{2-N^2+N^2\*z\over N\*(N^2-1)}(\gvb^2+\gab^2){1\over 1+z}\*\nn\\
&&\Bigg\{4\*(s+\mz^2)+(s+4\*\mz^2)\*z+(5\*s+2\*\mz^2)\*z^2\nn\\
&-&(1+z)\*(-2\*s+(3\*s+2\*\mz^2)\*z)\*\ln\left({s\over\mz^2}\right)\nn\\
&-&{s\*(1+z)+2\*\mz^2\over s\*(1+z)}\*(6\*s+4\*\mz^2-(s-2\*\mz^2)\*z+s\*z^2)\ln\left(1-{t\over\mz^2}\right)\nn\\
&-&{2\*(s+\mz^2)^2\*(1+z)\*(4-z+z^2)\over s\*(1-z)}\nn\\
&\times&
\left[\Li2{-{s\over\mz^2}}+\ln\left({s\over\mz^2}\right)\*\ln\left(1+{s\over\mz^2}\right)\right]\nn\\
&+&{2\over s\*(1-z)}\*\Big(5\*s^2+12\*s\*\mz^2+8\*\mz^4+2\*s\*(s+2\*\mz^2)\*z+s^2\*z^2\Big)\nn\\
&\times&\left[\Li2{t\over\mz^2}+\ln\left(s\over\mz^2\right)\*\ln\left(1-{t\over\mz^2}\right)\right]\Bigg\}+(z\ra-z),
\label{eq:gg-bbb-Z-box}
\end{eqnarray}
\begin{eqnarray}
\sumqn|\M^{\Box_W(gg\ra\bbb)}|^2 &=& 4\*\pi\alpha_s^2\alpha\*{1\over s}\*{2-N^2+N^2\*z\over N\*(N^2-1)}\*\gw^2\*{1\over 1+z}\*\nn\\
&&\Bigg\{s\*z\*(1+z)+{2\over \mt^2-\mw^2}\*\Big(2\*(s-\mt^2+\mw^2)\nn\\
&-&2\*(\mt^2-\mw^2)\*z+(2\*s-\mt^2+\mw^2)\*z^2\Big)\*\Big(\Amw-\Amt\Big)\nn\\
&-&(1+z)\*\Big(2\*s-(3\*s-2\*\mt^2+2\*\mw^2)\*z\Big)\*\Bosmtmt\nn\\
&+&\Big(6\*s-4\*\mt^2+4\*\mw^2-(s+2\*mt^2-2\*\mw^2)\*z+s\*z^2\Big)\*\Botmtmw\nn\\
&-&{1\over1-z}\*\Big(5\*s^2+4\*s\*(3\*\mw^2-2\*\mt^2)+8\*(\mt^2-\mw^2)^2\nn\\
&+&2\*s\*(s-2\*\mt^2+2\*\mw^2)\*z+s^2\*z^2\Big)\nn\\
&\times&\*\Big(\Cosmtmtmt-(1+z)\*\Cotmtmtmw\Big)\nn\\
&-&{1\over1-z}\*\Big[s\*(3\*s-4\*\mt^2+4\*\mw^2)\nn\\
&+&\Big(2\*s\*(2\*s-3\*\mt^2+4\*\mw^2)+6\*(\mt^2-\mw^2)^2\Big)\*z \nn\\
&-&s^2\*z^2+2\*\Big(s\*(s-\mt^2+2\*\mw^2)+(\mt^2-\mw^2)^2\Big)\*z^3\Big]\nn\\
&\times&\Cosmtmwmt\nn\\
&-&{1\over2\*(1-z)}\*\Big[5\*s^3+22\*s^2\*\mw^2-10\*s^2\*\mt^2\nn\\
&+&16\*s\*(\mt^4+2\*\mw^4-3\*\mt^2\*\mw^2)-16\*(\mt^2-\mw^2)^3\nn\\
&+&s\*\Big(7\*s^2-16\*s\*\mt^2+20\*s\*\mw^2+16\*(\mt^2-\mw^2)^2\Big)\*z\nn\\
&+&s\*\Big(3\*s^2+6\*s\*\mw^2-14\*s\*\mt^2+8\*\mt^2\*(\mt^2-\mw^2)\Big)\*z^2+s^3\*z^3\Big]\*\Dotw\Bigg\}\nn\\
&+&(z\ra-z),
\end{eqnarray}
\begin{eqnarray}
\sumqn|\M^{\Box_\phi(gg\ra\bbb)}|^2 &=& 2\*\pi\*\alpha_s^2\alpha{1\over s}\*{2-N^2+N^2\*z\over N\*(N^2-1)}\*\gw^2\*{\mt^2\over\mw^2}\*{1\over 1+z}\*\nn\\
&&\Bigg\{s\*z\*(1+z)+2\*\Big(2\*(1+z)+z^2\Big)\*\Big(\Amt-\Amw\Big)\nn\\
&-&(1+z)\*\Big(2\*s+(s+2\*\mt^2-2\*\mw^2)\*z\Big)\*\Bosmtmt\nn\\
&+&(2+z)\*\Big(s\*(1+z)-2\*(\mt^2-\mw^2)\Big)\*\Botmtmw\nn\\
&-&{1\over1-z}\*\Big(s^2+4\*s\*\mw^2+8\*(\mt^2-\mw^2)^2\nn\\
&+&2\*s\*(s-2\*\mt^2+2\*\mw^2)\*z+s^2\*z^2\Big)\nn\\
&\times&\*\Big(\Cosmtmtmt-(1+z)\*\Cotmtmtmw\Big)\nn\\
&-&{1\over1-z}\*\Big[s\*(s-4\*(\mt^2-\mw^2))\nn\\
&+&2\*\Big(s^2-s\*\mt^2+2\*s\mw^2+3\*(\mt^2-\mw^2)^2\Big)\*z\nn\\
&+&s^2\*z^2+2\*(s\*\mt^2+(\mt^2-\mw^2)^2)\*z^3\Big]\*\Cosmtmwmt\nn\\
&-&{1\over2\*(1-z)}\*\Big[s^3+6\*s^2\*\mw^2-2\*s^2\*\mt^2\nn\\
&+&16\*s\*\mw^2\*(\mw^2-\mt^2)+16\*(\mw^2-\mt^2)^3\nn\\
&+&s\*\Big(3\*s^2+12\*s\*\mw^2-8\*s\*\mt^2+16\*(\mt^2-\mw^2)^2\Big)\*z\nn\\
&+&s\*\Big(3\*s\*(s-2\*\mt^2+2\*\mw^2)+8\*\mt^2\*(\mt^2-\mw^2)\Big)\*z^2+s^3\*z^3\Big]\*\Dotw\Bigg\}\nn\\
&+&(z\ra-z).
\end{eqnarray}
%\bibliography{literatur}

\begin{thebibliography}{10}
%\cite{Ellis:1985er}
%\cite{Nason:1989zy}
\bibitem{Nason:1989zy}
  P.~Nason, S.~Dawson and R.~K.~Ellis,
  %``The One Particle Inclusive Differential Cross-Section for Heavy Quark
  %Production in Hadronic Collisions,''
  Nucl.\ Phys.\  B {\bf 327}, 49 (1989)
  [Erratum-ibid.\  B {\bf 335}, 260 (1990)].
  %%CITATION = NUPHA,B327,49;%%

%\cite{Beenakker:1988bq}
\bibitem{Beenakker:1988bq}
  W.~Beenakker, H.~Kuijf, W.~L.~van Neerven and J.~Smith,
  %``QCD Corrections to Heavy Quark Production in p anti-p Collisions,''
  Phys.\ Rev.\  D {\bf 40} (1989) 54.
  %%CITATION = PHRVA,D40,54;%%

\bibitem{Ellis:1985er}
  R.~K.~Ellis and J.~C.~Sexton,
  %``QCD Radiative Corrections To Parton Parton Scattering,''
  Nucl.\ Phys.\  B {\bf 269} (1986) 445.
  %%CITATION = NUPHA,B269,445;%%


%\cite{Giele:1993dj}
\bibitem{Giele:1993dj}
  W.~T.~Giele, E.~W.~N.~Glover and D.~A.~Kosower,
  %``Higher Order Corrections To Jet Cross-Sections In Hadron Colliders,''
  Nucl.\ Phys.\  B {\bf 403}, 633 (1993)
  [arXiv:hep-ph/9302225].
  %%CITATION = NUPHA,B403,633;%%

%\cite{Kuhn:2005az}
\bibitem{Kuhn:2005az}
  J.~H.~K\"uhn, A.~Kulesza, S.~Pozzorini and M.~Schulze,
  %``One-loop weak corrections to hadronic production of Z bosons at large
  %transverse momenta,''
  Nucl.\ Phys.\  B {\bf 727}, 368 (2005)
  [arXiv:hep-ph/0507178].
  %%CITATION = NUPHA,B727,368;%%

%\cite{Kuhn:2005gv}
\bibitem{Kuhn:2005gv}
  J.~H.~K\"uhn, A.~Kulesza, S.~Pozzorini and M.~Schulze,
  %``Electroweak corrections to hadronic photon production at large  transverse
  %momenta,''
  JHEP {\bf 0603}, 059 (2006)
  [arXiv:hep-ph/0508253].
  %%CITATION = JHEPA,0603,059;%%

%\cite{Kuhn:2007cv}
\bibitem{Kuhn:2007cv}
  J.~H.~K\"uhn, A.~Kulesza, S.~Pozzorini and M.~Schulze,
  %``Electroweak corrections to hadronic production of W bosons at large
  %transverse momenta,''
  Nucl.\ Phys.\  B {\bf 797}, 27 (2008)
  [arXiv:0708.0476 [hep-ph]].
  %%CITATION = NUPHA,B797,27;%%

%\cite{Moretti:2006nf}
\bibitem{Moretti:2006nf}
  S.~Moretti, M.~R.~Nolten and D.~A.~Ross,
  %``Weak corrections to gluon-induced top-antitop hadro-production,''
  Phys.\ Lett.\  B {\bf 639}, 513 (2006)
  [Erratum-ibid.\  B {\bf 660}, 607 (2008)]
  [arXiv:hep-ph/0603083].
  %%CITATION = PHLTA,B639,513;%%

%\cite{Moretti:2006ea}
\bibitem{Moretti:2006ea}
  S.~Moretti, M.~R.~Nolten and D.~A.~Ross,
  %``Weak corrections to four-parton processes,''
  Nucl.\ Phys.\  B {\bf 759}, 50 (2006)
  [arXiv:hep-ph/0606201].
  %%CITATION = NUPHA,B759,50;%%

%\cite{Beenakker:1993yr}
\bibitem{Beenakker:1993yr}
  W.~Beenakker, A.~Denner, W.~Hollik, R.~Mertig, T.~Sack and D.~Wackeroth,
  %``Electroweak one loop contributions to top pair production in hadron
  %colliders,''
  Nucl.\ Phys.\  B {\bf 411} (1994) 343.
  %%CITATION = NUPHA,B411,343;%%

%\cite{Kuhn:2005it}
\bibitem{Kuhn:2005it}
  J.~H.~K\"uhn, A.~Scharf and P.~Uwer,
  %``Electroweak corrections to top-quark pair production in quark-antiquark
  %annihilation,''
  Eur.\ Phys.\ J.\  C {\bf 45} (2006) 139
  [arXiv:hep-ph/0508092].
  %%CITATION = EPHJA,C45,139;%%

%\cite{Kuhn:2006vh}
\bibitem{Kuhn:2006vh}
  J.~H.~K\"uhn, A.~Scharf and P.~Uwer,
  %``Electroweak effects in top-quark pair production at hadron colliders,''
  Eur.\ Phys.\ J.\  C {\bf 51} (2007) 37
  [arXiv:hep-ph/0610335].
  %%CITATION = EPHJA,C51,37;%%

%\cite{Bernreuther:2006vg}
\bibitem{Bernreuther:2006vg}
  W.~Bernreuther, M.~Fuecker and Z.~G.~Si,
  %``Weak interaction corrections to hadronic top quark pair production,''
  Phys.\ Rev.\  D {\bf 74}, 113005 (2006)
  [arXiv:hep-ph/0610334].
  %%CITATION = PHRVA,D74,113005;%%

%\cite{Kuhn:1999de}
\bibitem{Kuhn:1999de}
  J.~H.~Kuhn and A.~A.~Penin,
  %``Sudakov logarithms in electroweak processes,''
  arXiv:hep-ph/9906545.
  %%CITATION = HEP-PH/9906545;%%

%\cite{Kuhn:1999nn}
\bibitem{Kuhn:1999nn}
  J.~H.~Kuhn, A.~A.~Penin and V.~A.~Smirnov,
  %``Summing up subleading Sudakov logarithms,''
  Eur.\ Phys.\ J.\  C {\bf 17}, 97 (2000)
  [arXiv:hep-ph/9912503].
  %%CITATION = EPHJA,C17,97;%%

%\cite{Kuhn:2001hz}
\bibitem{Kuhn:2001hz}
  J.~H.~Kuhn, S.~Moch, A.~A.~Penin and V.~A.~Smirnov,
  %``Next-to-next-to-leading logarithms in four-fermion electroweak  processes
  %at high energy,''
  Nucl.\ Phys.\  B {\bf 616}, 286 (2001)
  [Erratum-ibid.\  B {\bf 648}, 455 (2003)]
  [arXiv:hep-ph/0106298].
  %%CITATION = NUPHA,B616,286;%%

%\cite{Feucht:2004rp}
\bibitem{Feucht:2004rp}
  B.~Feucht, J.~H.~Kuhn, A.~A.~Penin and V.~A.~Smirnov,
  %``Two-loop Sudakov form factor in a theory with mass gap,''
  Phys.\ Rev.\ Lett.\  {\bf 93}, 101802 (2004)
  [arXiv:hep-ph/0404082].
  %%CITATION = PRLTA,93,101802;%%

%\cite{Jantzen:2005az}
\bibitem{Jantzen:2005az}
  B.~Jantzen, J.~H.~Kuhn, A.~A.~Penin and V.~A.~Smirnov,
  %``Two-loop electroweak logarithms in four-fermion processes at high
  %energy,''
  Nucl.\ Phys.\  B {\bf 731}, 188 (2005)
  [Erratum-ibid.\  B {\bf 752}, 327 (2006)]
  [arXiv:hep-ph/0509157].
  %%CITATION = NUPHA,B731,188;%%
%%%%%%%%%%%%%%%%%%%%%%%%
%\cite{Beccaria:1998qe}
\bibitem{Bec}
  M.~Beccaria, G.~Montagna, F.~Piccinini, F.~M.~Renard and C.~Verzegnassi,
  %``Rising bosonic electroweak virtual effects at high energy e+ e-
  %colliders,''
  Phys.\ Rev.\  D {\bf 58} (1998) 093014
  [arXiv:hep-ph/9805250].
  %%CITATION = PHRVA,D58,093014;%%

%\cite{Ciafaloni:1998xg}
\bibitem{CiaCom}
  P.~Ciafaloni and D.~Comelli,
  %``Sudakov enhancement of electroweak corrections,''
  Phys.\ Lett.\  B {\bf 446} (1999) 278
  [arXiv:hep-ph/9809321].
  %%CITATION = PHLTA,B446,278;%%

%\cite{Fadin:1999bq}
\bibitem{Fad}
  V.~S.~Fadin, L.~N.~Lipatov, A.~D.~Martin and M.~Melles,
  %``Resummation of double logarithms in electroweak high energy 
processes,''
  Phys.\ Rev.\  D {\bf 61} (2000) 094002
  [arXiv:hep-ph/9910338].
  %%CITATION = PHRVA,D61,094002;%%
\bibitem{Bec2}   M. Beccaria {\it et al.},  {Phys. Rev.} {D 61} (2000)
                 011301; {D 61} (2000)  073005.
\bibitem{DenPoz} 
%\cite{Denner:2000jv}
%\bibitem{Denner:2000jv}
  A.~Denner and S.~Pozzorini,
  %``One-loop leading logarithms in electroweak radiative corrections. I:
  %Results,''
  Eur.\ Phys.\ J.\  C {\bf 18} (2001) 461
  [arXiv:hep-ph/0010201].
  %%CITATION = EPHJA,C18,461;%%
%\cite{Denner:2001gw}
%\bibitem{Denner:2001gw}
  A.~Denner and S.~Pozzorini,
  %``One-loop leading logarithms in electroweak radiative corrections. II:
  %Factorization of collinear singularities,''
  Eur.\ Phys.\ J.\  C {\bf 21} (2001) 63
  [arXiv:hep-ph/0104127].
  %%CITATION = EPHJA,C21,63;%%


%\cite{Maina:2003is}
\bibitem{Maina:2003is}
  E.~Maina, S.~Moretti, M.~R.~Nolten and D.~A.~Ross,
  %``One-loop weak corrections to the b anti-b cross section at TeV energy
  %hadron colliders,''
  Phys.\ Lett.\  B {\bf 570} (2003) 205
  [arXiv:hep-ph/0307021].
  %%CITATION = PHLTA,B570,205;%%


%\cite{Passarino:1978jh}
\bibitem{Passarino:1978jh}
  G.~Passarino and M.~J.~G.~Veltman,
  %``One Loop Corrections For E+ E- Annihilation Into Mu+ Mu- In The Weinberg
  %Model,''
  Nucl.\ Phys.\  B {\bf 160} (1979) 151.
  %%CITATION = NUPHA,B160,151;%%

%\cite{vanOldenborgh:1990yc}
\bibitem{vanOldenborgh:1990yc}
  G.~J.~van Oldenborgh,
  %``FF: A Package to evaluate one loop Feynman diagrams,''
  Comput.\ Phys.\ Commun.\  {\bf 66} (1991) 1.
  %%CITATION = CPHCB,66,1;%%

%\cite{Denner:1991kt}
\bibitem{Denner:1991kt}
  A.~Denner,
  %``Techniques for calculation of electroweak radiative corrections at the one
  %loop level and results for W physics at LEP-200,''
  Fortsch.\ Phys.\  {\bf 41} (1993) 307
  [arXiv:0709.1075 [hep-ph]].
  %%CITATION = FPYKA,41,307;%%

%\cite{Kuhn:1987nh}
\bibitem{Kuhn:1987nh}
  J.~H.~Kuhn and R.~G.~Stuart,
  %``COMPACT FORM FOR THE CONTRIBUTION FROM Z - gamma BOX DIAGRAMS IN e+ e-
  %ANNIHILATION,''
  Phys.\ Lett.\  B {\bf 200}, 360 (1988).
  %%CITATION = PHLTA,B200,360;%%

%\cite{Jadach:1987ws}
\bibitem{Jadach:1987ws}
  S.~Jadach, J.~H.~Kuhn, R.~G.~Stuart and Z.~Was,
  %``QCD AND QED CORRECTIONS TO THE LONGITUDINAL POLARIZATION ASYMMETRY,''
  Z.\ Phys.\  C {\bf 38}, 609 (1988)
  [Erratum-ibid.\  C {\bf 45}, 528 (1990)].
  %%CITATION = ZEPYA,C38,609;%%

%\cite{Catani:1996vz}
\bibitem{Catani:1996vz}
  S.~Catani and M.~H.~Seymour,
  %``A general algorithm for calculating jet cross sections in NLO QCD,''
  Nucl.\ Phys.\  B {\bf 485} (1997) 291
  [Erratum-ibid.\  B {\bf 510} (1998) 503]
  [arXiv:hep-ph/9605323].
  %%CITATION = NUPHA,B485,291;%%


\end{thebibliography}
\newpage
\newcommand{\zp}{Z. Phys. }\def\as{\alpha_s }\newcommand{\prd}{Phys. Rev.
  }\newcommand{\pr}{Phys. Rev. }\newcommand{\prl}{Phys. Rev. Lett.
  }\newcommand{\npb}{Nucl. Phys. }\newcommand{\psnp}{Nucl. Phys. B (Proc.
  Suppl.) }\newcommand{\pl}{Phys. Lett. }\newcommand{\ap}{Ann. Phys.
  }\newcommand{\cmp}{Commun. Math. Phys. }\newcommand{\prep}{Phys. Rep.
  }\newcommand{\jmp}{J. Math. Phys. }\newcommand{\rmp}{Rev. Mod. Phys. }

\end{document}